\documentclass[rmp, aps]{revtex4}
\usepackage{graphicx}
\usepackage{amsmath}
\usepackage{natbib}
\usepackage{verbatim}
\usepackage{hyperref}

\newcommand{\fit}{F}  
\newcommand{\fitcoeff}{f}  
\newcommand{\acoeff}{a}  
\newcommand{\fitLEcoeff}{\hat{\fitcoeff}}  

\newcommand{\mfit}{\langle \fit \rangle} 
\newcommand{\mOfit}{\bar{\fit}} 
\newcommand{\sscale}{\bar{\sigma}}
\newcommand{\gt}{g}  
\newcommand{\locus}{s}
\newcommand{\rec}{r}
\newcommand{\xo}{c}
\newcommand{\mut}{\mu}
\newcommand{\mutop}{M}

\newcommand{\gtm}{\gt^{(m)}}
\newcommand{\gtf}{\gt^{(f)}}

\newcommand{\sigA}{\sigma^2_A}
\newcommand{\sigI}{\sigma^2_I}

\newcommand{\EQ}[1]{Eq.~(\ref{eq:#1})}

\begin{document}
\title{Statistical Genetics and Evolution of Quantitative Traits}

\author{Richard~A.~Neher${}^{\dagger{}^*}$}
\author{Boris~I.~Shraiman${}^{\dagger \ddagger}$}
\affiliation{${}^{\dagger}$Kavli Institute for Theoretical Physics}
\affiliation{${}^{\ddagger}$Department of Physics, University of California,
Santa Barbara, CA, 93106} 
\affiliation{${}^*$Max-Planck-Institute for Developmental Biology, T\"ubingen,
Germany} 

\date{\today}

\begin{abstract} 
The distribution and heritability of many traits depends
on numerous loci in the genome. In general, the
astronomical number of possible genotypes makes the system with large numbers of
loci difficult to describe. Multilocus evolution, however, greatly simplifies in
the limit of weak selection and frequent recombination. In this limit,
populations rapidly reach Quasi-Linkage Equilibrium (QLE) in which the dynamics
of the full genotype distribution, including correlations between alleles at
different loci, can be parameterized by the allele frequencies. This review
provides a simplified exposition of the concept and mathematics of QLE which is
central to the statistical description of genotypes in sexual populations. We
show how key results of Quantitative Genetics such as the generalized Fisher's
``Fundamental Theorem'', along with Wright's Adaptive Landscape, emerge within
QLE from the dynamics of the genotype distribution. We then discuss under what
circumstances QLE is applicable, and what the breakdown of QLE implies for the
population structure and the dynamics of selection. Understanding of the
fundamental aspects of multilocus evolution obtained through simplified models
may be helpful in providing conceptual and computational tools to address the
challenges arising in the studies of complex quantitative phenotypes of practical
interest.
\end{abstract}

\maketitle
\tableofcontents

\section{Introduction} R.A. Fisher's celebrated ``Fundamental theorem of natural
selection", relating the rate of change in the average fitness to the genetic
variance in fitness, occupies a place in Population Genetics
similar to Newton's ``$F=ma$" in Physics. Yet conceptually Fisher's law
and the whole subject of ``Quantitative Genetics"
\citep{Lynch:1998p8721,ScottFalconer:1996p10402}, which studies the response of
quantitative traits to selection, is closer to Thermodynamics.
Thermodynamics is a phenomenological description of readily measurable physical
properties (e.g.~average energy or pressure) of a large ensemble of molecules.
Quantitative Genetics is a phenomenological description of readily observable
phenotypic traits of a population. Thermodynamics takes macroscopic averages over
the random motion of individual molecules in thermal equilibrium. Quantitative
Genetics similarly focuses on the behavior of population-wide averages (and
variances) over many genetically diverse individuals. The genetic composition
of the population is governed by natural selection and random drift along with
recombination and mutation, all acting on individuals. The phenotype distribution
is related to the genotype distribution by the largely unknown
genotype-to-phenotype map, which is further obscured by environmental effects
which can cause phenotypic variation even between genetically identical individuals.
Yet deterministic laws of Thermodynamics emerge despite the complexity and chaos
of molecular motion. In fact they emerge {\it thanks to} the microscopic
complexity and chaos and  are made possible by the extensive self-averaging that
dominates macroscopic behavior of physical matter. Similarly, simple laws of quantitative population genetics
emerge when phenotypic traits depend on large numbers of polymorphic genetic loci.

\begin{figure}[tp]
\begin{center}
  \includegraphics[width=8.5cm]{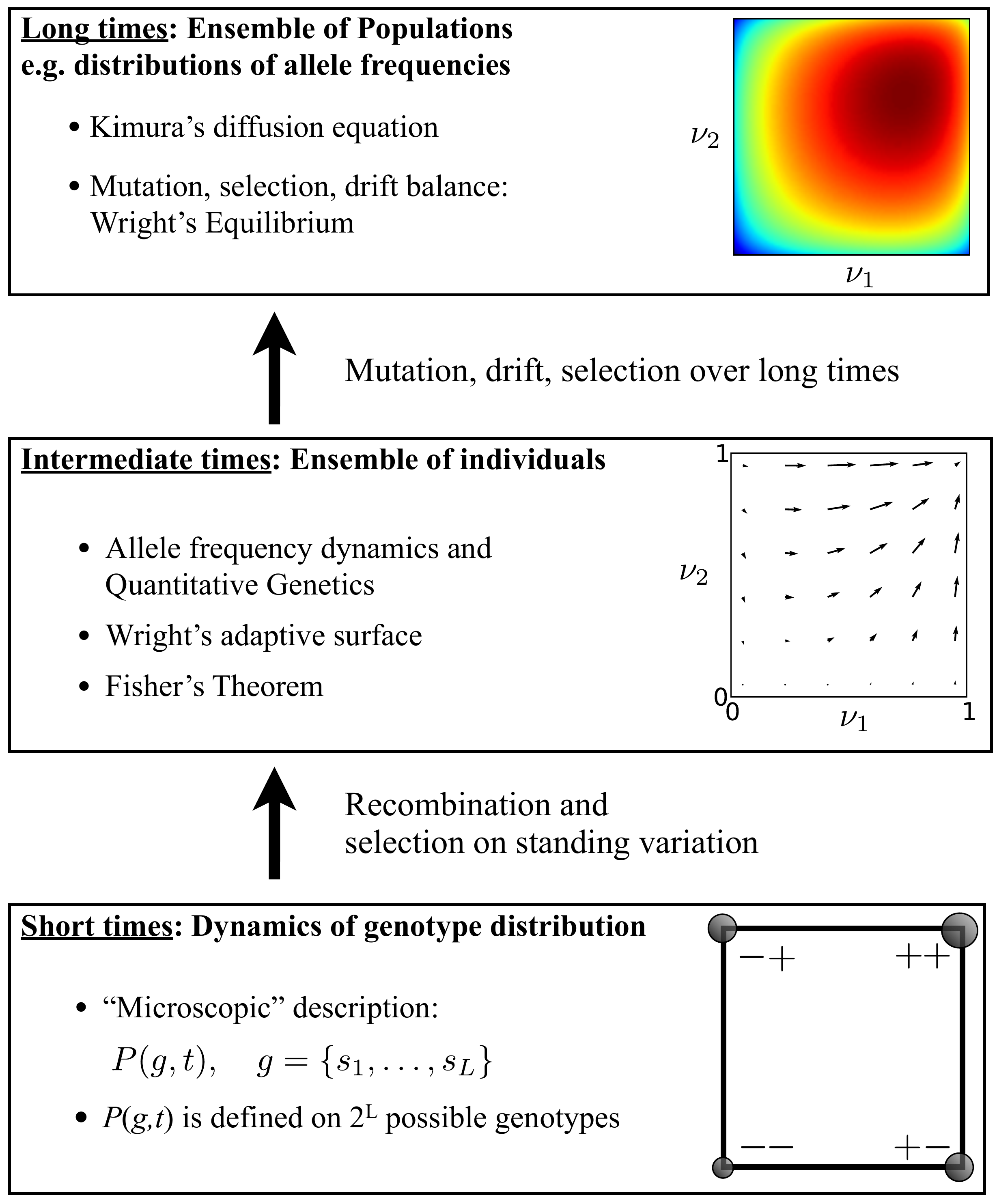}
  \caption[labelInTOC]{Time scales in sexual populations. A population is described by the distribution of 
  $2^L$ genotypes, on which selection, recombination and mutation acts. In sexual populations, 
  mating and recombination is the fastest process, so that different loci are
  only weakly correlated (close to linkage
  equilibrium) and its dynamics can be approximately described via $L$ allele
  frequencies -- a number much smaller  than the $2^L$  possible genotypes that
  would have to be tracked otherwise. Allele frequencies change slowly and means of quantitative traits 
  follow the laws of quantitative genetics. Over the much (much) longer time-scale of $\mu^{-1}$, allele frequencies themselves 
  tend to an equilibrium between selection, mutation, and genetic drift (assuming a constant environment). }
  \label{fig:sketch}
\end{center}
\end{figure}

While the analogy between Quantitative Genetics (QG) and Thermodynamics is most
appealing and has been noted by many including R.A. Fisher himself
\citep{Fisher_1930} - see
\citet{Iwasa:1988p29290,Sella:2005p28955,Barton:2009p21576} for recent work -
fundamental issues such as the lack of energy-like conserved quantity in
population genetics impede direct transcription of thermodynamic laws to QG.
Instead, the analogy must be pursued as an approach to the construction of a
coarse-grained phenomenological theory bridging the gap between ensemble averaged
(read population averaged) observables and the hidden micro-scale (read
individual genotype) dynamics. One must  be careful to define an averaging
ensemble that equilibrates on the time-scale of the observation, e.g.~the
response to selection in QG. In particular, as we illustrate in
Figure~\ref{fig:sketch}, dynamics in sexually reproducing populations are 
characterized by two widely different time-scales: 1) mating and
recombination reshuffle the polymorphic loci, allowing exploration of the space
of genotypes on a  short time scale and 2) mutation and population drift  control
genetic variation on much longer time scales, often long enough to render the
ensemble meaningless.

The bridge between the dynamics of the genotype distribution and the coarse
grained, QG-type, description is built on understanding multi-locus evolution.
Our review will focus on the intermediate time scale in the above
mentioned hierarchy. We will show how the genotype distribution $P(\gt,t)$ can be
parameterized by slowly varying allele frequencies, while mating and
recombination lead to rapid equilibration of $P(\gt,t)$ given a set of allele
frequencies. In this ensemble trait distributions are determined by allele
frequencies and the dynamics of trait averages can be expressed in terms of the
dynamics of allele frequencies. This in turn gives rise to the familiar laws of
quantitative genetics in terms of additive variances and covariances. In this
sense, a statistical multi-locus theory plays the role of Statistical Mechanics,
which explains how the deterministic laws of thermodynamics emerge from the
erratic motion of many microscopic particles. Hence the subject of the present
review should be thought of as ``Statistical Genetics" - a term introduced in a
closely related context by  \citet{Wright:1942p11795}.

Classical Quantitative Genetics \citep{ScottFalconer:1996p10402} is based on the
assumption that genotypes are random re-assortments of alleles, each occurring
with a certain frequency. This absence of correlations between alleles at
different loci is termed ``linkage equilibrium'', implying that recombination
(breaking linkage) has relaxed correlations between loci. 
This drastic simplification has earned QG a derogatory epithet of ``beanbag
genetics" from the pen of Ernst Mayr \citep{Mayr:1963} (see however \citet{Haldane:1964} in defense of beanbag
genetics). Yet in the present review we shall see that the key phenomenological laws
of QG extend beyond the assumption of linkage equilibrium. This understanding
emerges from the studies of multilocus selection which began with  two
alleles/two loci systems
\citep{KIMURA:1956p33458,Lewontin:1960p20638,Karlin:1969p33606}.
\citet{Kimura:1965p3008} showed that a two locus system tends towards a
state where allele frequencies change slowly and correlations are small and
steady. He termed this state Quasi-Linkage Equilibrium (QLE), which is the
subject of this review. Subsequently, several comprehensive treatments of
multilocus evolution were developed \citep{Christiansen:1990p32416,Barton:1991p2659,BURGER:1991p2662,
NAGYLAKI:1993p2698,PrugelBennett:1994p11485,Baake:2001p33421} (for a monograph
see \citet{Buerger2000}) with \citet{Barton:1991p2659} and \citet{NAGYLAKI:1993p2698}, in particular,
generalizing and justifying the QLE approximation  in multi-locus systems.

In addition to the study of generic behavior of systems with a very large
number of loci, explicit multi-locus modeling of smaller systems has been used to
study the evolution of recombination \citep{Barton:1995p2811,Roze:2006p2922} and
patterns of genetic variation produced by positive selection
\citep{Stephan:2006p12174}. Recent work produced interesting 
examples \citep{Weinreich:2006p15529,deVisser:2009p37564} 
of empirically determined fitness landscapes with five or more loci.
The dynamics of populations on these landscapes can be studied in
laboratory experiments and comparison to theoretical models is possible
\citep{deVisser:2009p37564}. Quantitative understanding of multi-locus evolution is
also essential when studying the emergence of drug resistance in HIV,
which often depends on several interacting loci in a recombining population
\citep{Kellam:1994p27753,Nora:2007p20790,Bretscher:2004p24459}.

Our discussion of the multilocus selection problem will follow the Barton-Turelli
course of making and keeping it simple
\citep{Barton:1991p2659,Turelli:1994p2652}, by attempting to make it simpler
still. We will define a streamlined conceptual and analytic framework which will not
only reproduce classic results, but also readily generate some new extensions.
To accomplish this we will formulate and analyze a ``minimal model" of
multilocus evolution: continuous time selection in a haploid model with a general
(epistatic) fitness function of $L$ loci. We shall show how the Generalized
Fisher Theorem and other results of quantitative genetics follow from a
straightforward cumulant perturbation theory similar to that used extensively
(for high temperature expansions) in Statistical Physics \citep{mcquarrie:1973}.
The perturbative regime corresponds to Kimura's Quasi-Linkage Equilibrium (QLE).
Using this formulation of QLE, we present systematic generalizations of QG
results and of (Kimura's) diffusion theory, typically formulated in complete linkage
equilibrium, to include weak correlations between loci. We also discuss how QLE
breaks down when the ratio of characteristic strength of selection to the rate of
recombination exceeds a critical value that depends on the strength of epistasis.
While the QLE regime corresponds to selection of individual alleles based on
their effect on fitness averaged over genetic backgrounds, the breakdown of QLE
follows the appearance of strong correlations between alleles at different loci
and represents a transition to effective selection of genotypes. In the
Discussion section we shall connect the transition from ``allele selection" to
``genotype selection" to the closely related spin-glass transition
(modeling the behavior of disordered magnets) studied in Statistical
Physics \citep{Mezard:1987p21192}. We shall also discuss its implications for
Quantitative Genetics.

\section{Relating quantitative traits and genotypes.} 
\label{sec:geno_pheno}
Let us focus on the
fitness which is the most important example of ``quantitative trait", although
everything we shall say about ``fitness landscapes" in this section, applies
directly to any quantitative phenotype.  A fitness landscape is a metaphor for a
map from the high dimensional space of genotypes to expected reproductive
success. While the map itself is unambiguous, several different ways of
parameterizing fitness landscapes with alleles and groups of alleles have been proposed
\citep{Weinberger:1991p17543,Barton:1991p2659,Hansen:2001p4472,Hansen:2006p9516}.

Consider a haploid genome of $L$ loci with two alleles each, such that a genotype
is uniquely characterized by $L$ binary variables $\gt=\{\locus_1, \ldots,
\locus_L\}$. We choose $\locus_i\in\{-1, 1\}$, $i=1,\dots,L$ instead of 
$\locus_i\in\{0, 1\}$ (more commonly used in population genetics literature), since the symmetric choice simplifies the algebra
below. (The relation between representations can be
found in Appendix ~\ref{sec:notation}, a short glossary of
population genetics terminology is given in Appendix \ref{sec:glossary}.)
Functions of the genotype, e.g.~as the population distribution, fitness, or any other quantitative trait, 
live therefore on a $L$-dimensional hypercube. Any such
function on the hypercube can be decomposed into a sum of monomials in $\locus_i$
\begin{equation}
\label{eq:orthogonal_decomp}
\fit(\gt)=\mOfit+\sum_i
\fitcoeff_i\locus_i+\sum_{i<j}\fitcoeff_{ij}\locus_i\locus_j+\sum_{i<j<k}\fitcoeff_{ijk}\locus_i\locus_j\locus_k+\cdots,
\end{equation}
where the first sum represents independent contribution of $L$ single loci, the
second sum which runs over all $L(L-1)/2$ pairs of loci represents contribution
of pairs and the higher order terms account for the effect of each and
every possible subgroups of loci. The first order contribution $\fitcoeff_i$ defines 
the additive effect of locus
$i$ which is independent of all other loci
considered. Higher order terms which include locus $i$ define the genetic
background dependence of the effect of $\locus_i$ allele. Collectively, terms of
order higher than one represent genetic interactions also known as
``epistasis". The contribution of each locus or subgroup of
loci is determined by unbiased (i.e.~each genotype enters with weight $2^{-L}$)
averaging over the remainder of the genome: thus the coefficients are given by
\begin{equation}
\label{eq:coefficients}
\mOfit=2^{-L}\sum_{\gt}\fit(\gt), \quad
\fitcoeff_i=2^{-L}\sum_{\gt}\locus_i \fit(\gt), \quad
\fitcoeff_{ij}=2^{-L}\sum_{\gt}\locus_i\locus_j \fit(\gt), \ldots
\end{equation}
One easily convinces oneself that plugging \EQ{orthogonal_decomp} 
into the expressions in \EQ{coefficients} reduces to the desired coefficients. 
In total, there are $2^L$ coefficients $f^{(k)}_{i_1, \ldots, i_k}$, as it has to be
for an exact representation of a function on a hypercube. In fact, the coefficient of
the expansion of $\fit(\gt)$ into monomials is nothing but the Fourier transform of
the original function on the hypercube, which was used in the context of
genotype-fitness maps in
\citet{Weinberger:1991p17543,Stadler:1998p17511,Hordijk98amplitudespectra}. In
addition to this genetic contribution to the trait, the trait value of a given
individual will also depend on environmental (and epigenetic) factors which are not modeled here.

It proves very useful to define a ``density of states" $\rho
(F)=2^{-L}\sum_{\gt} \delta (F-\fit(\gt))$, where $\delta(F)$ is a Dirac
delta-function. The fraction of genotypes with fitness in the interval $[F,
F+\delta F]$ is then given by $\int_F^{F+\delta F} dF' \rho(F')$. Provided
$\fit(\gt)$ receives contributions of very many terms of similar magnitude in
Eq. \ref{eq:orthogonal_decomp}, the Central Limit Theorem will apply making the density
of states approximately Gaussian in shape. The width of this Gaussian is given by
the (square root of the) variance of $\fit(\gt)$ over the hypercube:
\begin{equation}
\label{eq:powerspectrum}
\sscale^2 =2^{-L}\sum_{\gt}  (\fit(\gt) - \bar \fit )^2 = \sum_i f_i^2 +
\sum_{i<j} f_{ij}^2 +  \sum_{i<j<k} f_{ijk}^2 +...
\end{equation}
This simple decomposition of variance is the equivalent of the Parseval's theorem
 for the Fourier transform. Note that this variance is an
intrinsic property of the fitness landscape completely independent of any population that
may be evolving on it. It should not be confused with the population variance
that we will discuss later. We shall use $\sscale$ as a measure of selection
strength.

The sums in Eq.~(\ref{eq:powerspectrum}) for $\bar{\sigma}$ can be interpreted as
the power spectrum of the $\fit(\gt)$. A falling or rising power spectrum gives rise
to qualitatively different landscapes: If most of the variation of the fitness
function were captured by the first order terms, the landscape would be smooth and simple. 
If higher order terms dominate
the fitness variance, the landscape is multi-peaked and rugged. The properties of
smooth versus rugged landscapes (parameterized in the manner of
\EQ{orthogonal_decomp} ) has been a subject of extensive study in statistical
physics as it relates to the theory of spin-glasses \citep{Mezard:1987p21192}. 
It is known that the key consequences of complexity of the general landscape,
appear already in the class of functions involving only pairwise interactions
(a.k.a.~the Sherrington-Kirkpatrick model \citep{Sherrington:1975p36835})
\citep{Mezard:1987p21192}. Here for simplicity we shall consider only pairwise
interactions. (An alternative instructive simplification would be to consider $\fit(\gt )$ defined
by a fixed random function on the hypercube, known in population genetics as the
``house-of-cards" model \citep{Kingman:1978p31199} or NK-models
\citet{Kauffman:1989p39602} and in physics as a ``random energy" model
\citep{Derrida:1981p30865}.)

Before moving on to population dynamics, it is instructive to discuss the
implication of the combinatorial explosion of higher order interactions: In
principle there are ${ L \choose k }$ interactions of order $k$, a number which
increases with $L$ as $L^k$. Hence increasing the number of loci without changing
the statistics of the coefficients would shift the power spectrum towards higher
order, making the function more rugged. It seems more likely that the
interactions are sparse with the number of ``partners" of a typical locus not
growing in proportion to the total number of loci: In particular one may posit
that each locus interacts with a finite number of other loci, independent of $L$
and set all other coefficients to zero. Unfortunately, despite some recent
progress \citep{Brem:2005p4975,Ehrenreich:2010p34021}, we still know very little
about generic structure of genotype-phenotype maps. One must also be aware of the
fact that, because of  selection, statistics of genetic interactions observed
among co-segregating polymorphisms within a breeding population may be quite
different from that for a random set of loci or a for polymorphisms created by
crossing two isolated populations \citep{Jinks:1966p12252}. Indeed, most
immediate evidence for epistasis is provided by the ``outcrossing depression":
suppression in the fitness of progeny issuing from a cross of diverged strains
\citep{Jinks:1966p12252,Seidel:2008p6677}.

\section{Dynamics of the genotype distribution.} 
\label{sec:GT_Dynamics}
Selection, mutation, and recombination operate on individuals and change the
distribution of genotypes, $P(\gt, t)$, in the population. The fitness $\fit(\gt)$ 
of a genotype $\gt$ is defined as the expected reproductive
success, i.e.~the rate at which the proportion of a genotype increases or
shrinks in the population due to (natural or artificial) selection. During the time interval $\Delta t$,
\emph{selection} changes the distribution of genotypes according to
\begin{equation}
\label{eq:selection}
P(\gt , t+\Delta t) = \frac{e^{\Delta t \fit(\gt)}}{\langle e^{\Delta t\fit}\rangle}
P(\gt , t) 
\end{equation}
where $\langle e^{\Delta t \fit}  \rangle =  \sum_{\gt} e^{\Delta t \fit(\gt)}
P(\gt , t)$ denotes the population average. The genetic diversity that selection
acts upon is due to \emph{mutations}, which change the genotype distribution as
follows
\begin{equation}
\label{eq:mutation}
P(\gt, t+\Delta t)  = P(\gt, t) + \Delta t\mut \sum_{i=1}^L
\left[P(\mutop_i\gt,t)-P(\gt,t)\right] \ .
\end{equation}
Here $\mutop_i \gt$ is a shorthand for genotype $\gt$ with $\locus_i$ replaced
by $-\locus_i$. Despite the importance of mutations for generating polymorphisms
and maintaining genetic diversity in the long run, the effect of mutation on the
dynamics of significantly polymorphic sites can be neglected if
mutation rates are much smaller than selection coefficients.

In addition to selection and mutation, the dynamics of the genotype distribution
in sexual  populations are driven by \emph{mating and recombination}. Gametes
are formed during meiosis crossing over homologous parental chromosomes. Assuming
random pairing of gametes and outcrossing with rate $\rec$, the genotype
distribution changes during recombination as follows:
\begin{equation}
\label{eq:recombination}
P(\gt, t+\Delta t) = (1-\Delta t\rec)P(\gt, t) +  \Delta
t\rec\sum_{\{\xi_i\}\{\locus'_i\}} C(\{\xi\})P(\gtm,t)P(\gtf,t) \ .
\end{equation}
The first term accounts for those individuals that did not outcross during the
$\Delta t$ time interval. In the event of outcrossing, a new genotype is formed
from genetic material of the mother with genotype $\gtm$ and a father with
genotype $\gtf$. The novel recombinant genotype $\gt$ inherits a subset of his
loci from the mother and the complement from the father, which in
\EQ{recombination} is described by the set of random variables $\{\xi_i\}$. If
$\xi_i=1$, gene $i$ is inherited from the mother, if $\xi_i=0$ from the father.
Using this notation, the maternal genotype is $\locus^{(m)}_i=\xi_i \locus_i +
(1-\xi_i)\locus'_i$ and equivalently the paternal genotype
$\locus^{(f)}_i=(1-\xi_i) \locus_i + \xi_i \locus'_i$. The part of the maternal
and paternal genome which is not passed on to the offspring, $\{\locus_i'\}$,  is
summed over. Each particular realizations of $\{\xi_i\}$, i.e.~a pattern of
crossovers, has probability $C(\{\xi\})$, which depends on the crossover rates
between different loci. In addition to the summation over all $\{\locus_i'\}$, we
have to sum over possible crossover patterns $\{\xi_i\}$.  A very similar
notation was used in \citet{Christiansen:1990p32416}.  While our presentation so far
was completely general, dealing with diploid genomes inflates the required
book-keeping as we proceed with the analysis. Since our goal is to present the
key effects and ideas in the simplest possible form, we shall from here on
restrict to considering only haploids, two of which recombine upon mating
producing a haploid offspring. Although this model is chosen for simplicity
sake, it is sufficient to describe diploids in the absence of dominance. It 
also describes haploid yeast going through
mating/sporulation/germination cycle or the population genetics of many RNA
viruses like HIV and Influenza. 

Provided selection is weak ($\Delta t  \fit(\gt) \ll 1$), we can use a
continuous time description of the dynamics.
\begin{equation}
\label{eq:genotype_dynamics}
\begin{split}
\frac{d}{dt} P(\gt ,t) = &(\fit(\gt)-\mfit) P(\gt ,t)  + \mut
\sum_{i=1}^L \left[P(\mutop_i\gt)-P(\gt)\right]\\
&+\rec\sum_{\{\xi_i\}\{\locus'_i\}} C(\{\xi\})\left[P(\gtm ,t)P(\gtf ,t)-P(\gt'
,t)P(\gt ,t)\right]
\end{split}
\end{equation}
This equation describes the dynamics of the genotype distribution in the limit
$N\to \infty$ where each genotype is sampled by enough individuals to neglect
sampling noise which would arise during reproduction. This stochastic component
to the dynamics of the genotype distribution is known as ``random genetic drift''.
We shall discuss random drift in Section \ref{sec:drift}. Our focus here will be 
on the interplay between selection and recombination, which dominates the
behavior of Eq.~(\ref{eq:genotype_dynamics}).

Instead of specifying $P(\gt,t)$ for every $\gt$, $P(\gt,t)$ can be parameterized
by its cumulants. The cumulants of first and second order are defined as $\chi_i
= \langle \locus_i \rangle$ and
$\chi_{ij}=\langle\locus_i\locus_j\rangle-\langle\locus_i\rangle \langle
\locus_j\rangle$, which are related to allele frequencies and pairwise linkage
disequilibria (see Table \ref{tab:translation}). In total there are $2^L-1$ cumulants,  with
higher order ones, $\chi_{ij\ldots k}$, more easily defined via the cumulant generating
function \cite{mcquarrie:1973}. However, only the first and second order
cumulants will be needed in the present context.

To obtain dynamical equations for $\chi_i$, we multiply
Eq.~(\ref{eq:genotype_dynamics}) by $\locus_i$ and sum over all possible
genotypes. One finds
\begin{equation}
\label{eq:general_allele_freq_dyn}
\dot{\chi}_i 
= \langle \locus_i [\fit(\gt)-\mfit]\rangle - 2\mu\langle \locus_i \rangle \ ,
\end{equation}
where we have used $\mutop_i \locus_i = -\locus_i$ and used the notation
$\dot{\chi}_i$ for total derivative with respect to time. The dynamics of
$\chi_i$ do not depend explicitly on the recombination rate, which is
intuitive since recombination does not create or destroy alleles. In order to evaluate
$\langle \locus_i \fit(\gt) \rangle$ in Eq.~(\ref{eq:general_allele_freq_dyn}) we
need to know higher order cumulants, i.e.~we are faced with a
hierarchy of cumulant equations.


In contrast to first order cumulants, the dynamics of higher order cumulants depend 
explicitly on recombination, which has the tendency to destroy associations between alleles 
and drives higher order cumulants to zero. To write down an equation for the
dynamics of the second order cumulants, $\chi_{ij} = \langle
s_i s_j \rangle -\chi_i\chi_j$, we have to evaluate ${d \over dt}\langle
s_i s_j \rangle$, which explicitly depends on recombination. Evaluating the
recombination term only, we find
\begin{equation}
\label{eq:general_chi_ij_freq_dyn}
\rec\sum_{\{\xi_i\}}C(\{\xi\})\sum_{\gt,\gt'}
\locus_i\locus_j  \left[P(\gtm ,t)P(\gtf ,t)-P(\gt' ,t)P(\gt
,t)\right]= -\rec\xo_{ij}\chi_{ij}
\end{equation}
where $c_{ij}$ is the probability that loci $i$ and $j$ derive from different parents: $ c_{ij} = \sum_{ \{ \xi \} } C(\{ \xi \}) [\xi_i (1-\xi_j)+(1-\xi_i)\xi_j ]$. 
To arrive at this result, we substituted $\locus_i = \xi_i
\locus_i^{(m)}+(1-\xi_i)\locus_i^{(f)}$ (analogously for $\locus_j$), and 
averaged over the maternal and paternal genomes. The second term evaluates simply to $\rec
\langle \locus_i\locus_j \rangle$. This result holds more
generally for central moments of the genotype distribution \citep{Barton:1991p2659}.
Together with selection and mutation, we find (for $i \neq j$)
\begin{equation}
\dot{\chi}_{ij} = \langle
(\locus_i-\chi_i)(\locus_j-\chi_j)(\fit(\gt)-\mfit)\rangle-4\mu\chi_{ij}-\rec\xo_{ij}
\chi_{ij}
\end{equation}
We see that selection drives $\chi_{ij}$ away from zero, while $\chi_{ij}$
relaxes though mutation and recombination.  In absence of selection $P(\gt ,t)$
tends to a steady state of ``linkage equilibrium" (LE) with vanishing cumulants $\chi_{ij}$ (for $i \neq j$) implying complete decorrelation
of alleles at different loci corresponding to factorization of the genotype distribution: $P_0(\gt )=\prod_{i=1}^L
p_i(s_i)$. It is easy to see that the recombination term in
Eq.~(\ref{eq:genotype_dynamics}) vanishes whenever $P(\gt) = P_0(\gt)$.
In Section \ref{sec:QLE}, we will, starting from $P_0(\gt)$, develop the
Quasi-Linkage-Equilibrium (QLE) approximation by systematically accounting for
small linkage disequilibria ($\chi_{ij}$).


\section{Trait distributions and the dynamics of population averages}
In most cases, $P(\gt, t)$ cannot be observed directly. Instead, the subject of
quantitative genetics are distributions of traits in the population. Trait
distributions can be obtained from genotype distributions by projection. The
probability of finding in the population an individual with fitness (or any other
trait) in the interval $[F, F+\Delta F ]$ is given by
\begin{equation}
\label{eq:projection}
p(F, t) = \sum_{\gt} \delta (F - F (\gt ) ) 
P(\gt , t) 
\end{equation}
where $\delta (F)$ is the Dirac delta-function ($\int dF \delta (F) =1$).
Applying this projection to Eq.~(\ref{eq:genotype_dynamics}) yields an
equation for the dynamics of the trait distribution. Before addressing the dynamics of
traits in sexual populations, it is
instructive to consider the dynamics of the fitness distribution $p(\fit,t)$ in
absence of mutation and recombination, in which case one obtains simply
\begin{equation}
\label{eq:contin_selection_F}
{ d \over dt } p(\fit , t) = [ \fit-\langle \fit \rangle]
p(\fit , t) 
\end{equation}
where $\langle \fit \rangle = \int d\fit \fit p( \fit ,t)$. Multiplying this equation by $\fit$ and integrating over $\fit$
(i.e.~the 1st moment of this equation)  yields Fisher's ``Fundamental Theorem" in the asexual case.
\begin{equation}
\label{eq:FFTNS}
{ d \over dt } \langle \fit \rangle= \langle [ \fit-\langle \fit \rangle]^2
\rangle = \sigma^2
\end{equation}
Evidently this is just the 1st in the hierarchy of infinitely many moment
equations that characterize the dynamics of $p(F,t)$ given explicitly by
Eq.~(\ref{eq:contin_selection_F}). The 2nd moment expresses the dynamics of $\sigma^2$
in terms of the 3rd moment, etc. This hierarchy of equations is not closed, yet
under certain conditions higher moments may be suppressed making the $\sigma^2$
a slowly varying function of time. One notes that
Eq.~(\ref{eq:contin_selection_F}) has a Gaussian traveling wave solution $p(\fit
,t)=C \exp[ - (\fit -vt)^2/2v ]$ with an arbitrary constant variance $\sigma^2 = v$ 
setting the rate of fitness growth ${d \over dt } \langle \fit \rangle=v$ in
agreement with Eq.~(\ref{eq:FFTNS}). A traveling wave with constant speed requires that 
genotypes with arbitrarily high fitness are populated with at least one
individual, which requires an infinitely large population with infinitely many polymorphic loci
with limits taken in this order. Otherwise genetic diversity disappears and
adaptation stalls. The evolution of the shape of the fitness distribution in
finite populations has been studied in the context of Genetic Algorithms by
\citet{PrugelBennett:1994p11485}. \citeauthor{PrugelBennett:1994p11485} study
the effect of selection and recombination on the cumulants of the fitness
distribution and observe how the fitness variation vanishes as the population
condenses into a local fitness maximum.
To prevent this condensation, new variation has to be constantly supplied by mutation. 
Quite generally $\sigma$ is determined by the balance generation of
genetic variation through mutations or recombination and its removal be
selection and drift, which requires careful stochastic treatment 
\citep{Tsimring:1996p19688,Rouzine:2003p33590,Rouzine:2005p17398,Desai:2007p954,Neher:2010p30641,Hallatschek:2011p39697}.

One can also consider the dynamics of an
arbitrary trait $G(\gt)$ different from fitness.
In analogy to Eq.~(\ref{eq:projection}), we can study the joint distribution,
$p(\fit, G, t)$ of this trait with fitness. The population average of the trait
obeys
\begin{equation}
\label{eq:price_equation}
{ d \over dt } \langle G \rangle = \langle G \fit \rangle - \langle G 
\rangle\langle \fit \rangle = Cov(F,G)_{P(\gt)}
\end{equation}
i.e.~its rate of change is given by its covariance with fitness
\citep{Price:1970p36449}. This statement is also known as the ``secondary theorem"
of natural selection \citep{Robertson:1966p36223}.

With mutation and recombination, the dynamics of trait means are no longer that
simple. To evaluate the mutation and recombination terms, we utilize the orthogonal expansion of the
fitness function in Eq.~(\ref{eq:orthogonal_decomp}). Restricting ourselves to
pairwise interactions, we can use Eqs.~(\ref{eq:general_allele_freq_dyn}) and
(\ref{eq:general_chi_ij_freq_dyn}) to obtain
\begin{equation}
\label{eq:general_mean_fitness_dyn}
\frac{d}{dt} \mfit = \sigma^2 -\mu \Delta_\mu -\rec \sum_{i<j}
\xo_{ij}\fitcoeff_{ij} \chi_{ij}\ ,
\end{equation}
where $\Delta_\mu$ is the average loss in fitness due to mutation. The latter
can be calculated by observing that each moment decays through mutation with rate $2\mu k$, where
$k$ is the order of the moment. 
\begin{equation}
\Delta_\mu = \mu \left[ 2\sum_i \fitcoeff_i\chi_i +
4\sum_{i<j}\fitcoeff_{ij}(\chi_i\chi_j+\chi_{ij}) +\cdots\right]
\end{equation}
Higher moments decay faster because they have a greater mutation target. 
The second term in Eq.~(\ref{eq:general_mean_fitness_dyn}) is the loss in
fitness through recombination, which reflects the tendency of recombination to
factorize the genotype distribution such that contributions like $\fitcoeff_{ij}\chi_{ij}$ to $\mfit$ decay with
rate $\rec \xo_{ij}$.  We will later see that the previous form of Fisher's theorem can be
recovered by a suitable definition of an \emph{additive} fitness variance. To do
so, however, we have to understand how the genotype distribution evolves under
selection and recombination.

\section{Beyond linkage equilibrium: Quasi Linkage Equilibrium} 
\label{sec:QLE}
We have already seen that without selection or without epistasis, $P(\gt )$
distribution will relax to a product of independent distributions at different
loci: the linkage equilibrium state. Next we shall account for the correlations
between loci induced by selection. For simplicity we shall omit the
mutational contribution, which we shall restore once we understand the basis of
Quasi Linkage Equilibrium (QLE). 

\subsection{QLE: A perturbation expansion at high recombination rates}
If selection  on the time scale of recombination is weak, i.e. $\sscale \ll
\rec$, the induced correlation is also weak and can be calculated using
perturbation theory \citep{Kimura:1965p3008,Barton:1991p2659}. To this end, we
parameterize the genotype distribution as follows
\begin{equation}
\label{eq:logP}
\log P(\gt , t) =  \Phi (t) +\sum_i \phi_i (t) \locus_i + \sum_{i<j}
\phi_{ij} (t) \locus_i \locus_j \ ,
\end{equation}
which is the already familiar Fourier representation of functions on the genotype
space. The factorized distribution $P_0 (\gt )$ would correspond to  the
coefficients, $\phi_{ij}$, of all multilocus contributions being zero. The
second order terms capture (to the leading order) the correlations induced by selection and (in the
limit under consideration) are assumed to be small.  The genotype independent
term ${\Phi}(t)$ is fixed by the normalization of the probability distribution.

\begin{equation}
\label{eq:partition }
e^{-{\Phi} (\{ \phi \} )} = \sum_g \exp \left [ \sum_i \phi_i  \locus_i +
\sum_{i<j} \phi_{ij}  \locus_i \locus_j \ +... \right ]
\end{equation}
and acts as the generator of the cumulants via
\begin{eqnarray}
\label{eq:gen_of_cumulants}
\chi_i &=& -{ \partial  {\Phi} \over \partial \phi_i } \ , \quad
\chi_{ij} = -{ \partial^2 {\Phi}\over \partial \phi_i \partial \phi_j }  \
\end{eqnarray}

The generating function $\Phi$ is evaluated perturbatively for small $\phi_{ij}$ in the 
Appendix \ref{sec:app_QLE} yielding 
\begin{eqnarray}
\label{eq:chi_to_phi}
\chi_i& \approx&
\tanh (\phi_i)+\sum_{j \neq i} \phi_{ij} (1-\tanh^2
(\phi_i))\tanh (\phi_j)\\ 
\chi_{ij} &\approx&
(1-\chi_i^2)(1-\chi_j^2)\phi_{ij} \; \;\; \;  \mathrm{for}\;  i \neq j \\
\chi_{ii} &=&1-\chi_i^2
\end{eqnarray}
which is correct to the leading order in $|\phi_{ij}|$.  The distribution given
by Eq. (\ref{eq:logP}) may be thought of as a maximum entropy distribution
constrained to have certain first and second order cumulants: Parameters
$\phi_i$ and $\phi_{ij}$ are the Lagrange multipliers that impose the constraints.

Let us rewrite Eq.~(\ref{eq:genotype_dynamics}) as an equation for the dynamics
of $\log P(\gt)$ which yields
\begin{equation}
\begin{split}
\label{eq:logPdt}
&\dot{{\Phi}}+\sum_i  \dot{\phi}_i \locus_i +
\sum_{i<j} \dot{\phi}_{ij}\locus_i\locus_j =\fit(\gt)-\mfit + \rec\sum_{\{\xi_i\}\{\locus'_i\}}
C(\{\xi\})P(\gt')\left[\frac{P(\gtm)P(\gtf)}{P(\gt)P(\gt')}-1\right] \\
& \approx \mOfit+\sum_i \fitcoeff_{i}\locus_i+\sum_{i<j}
\fitcoeff_{ij}\locus_{i}\locus_j +\rec
\sum_{i<j} c_{ij}\phi_{ij}\left[(\locus_i\langle
\locus_j\rangle+\langle \locus_i\rangle \locus_j)- (\locus_i \locus_j+\langle
\locus_i\locus_j\rangle) \right]
\end{split}
\end{equation}
where the recombination part has been evaluated approximately by expanding the
exponential that defines $P(\gt)$ (see Appendix \ref{sec:app_QLE}).
We can now collect terms with the same monomials in $\locus_i$ to obtain the
equations governing the time evolution of $\phi_i$ and $\phi_{ij}$:
\begin{eqnarray}
\label{eq:d_t_phi}
\dot{\phi}_i &  = & \fitcoeff_i + \rec \sum_{j\neq i} c_{ij}\phi_{ij}
\langle\locus_j\rangle \\
\dot{\phi}_{ij} & = & \fitcoeff_{ij} -\rec c_{ij} \phi_{ij} 
\end{eqnarray}
At large crossover rates $\rec c_{ij}$, the $\phi_{ij}$ rapidly approach a
steady-state $\phi_{ij} = \frac{\fitcoeff_{ij}}{\rec c_{ij}}$. This has to be
contrasted with the behavior in absence of recombination, in which case
$\phi_{ij}$ would grow linearly as $\fitcoeff_{ij} t$. Recombination prevents effective
selection on interactions. Instead, the higher order contributions to
fitness affect the dynamics of $\phi_i$ after averaging over possible
genetic backgrounds: Substituting the
steady state relation into the equation for $\phi_i$ yields 
\begin{equation}
\dot{\phi}_i = \fitcoeff_i + \sum_{j \neq i} \fitcoeff_{ij}\chi_j =
\fitLEcoeff_i\ .
\end{equation}
where we have defined $\fitLEcoeff_i = \fitcoeff_i + \sum_j \fitcoeff_{ij}\chi_j$
which is the effective strength of selection acting on locus $i$ in linkage
equilibrium. It is obtained from the general expression for $F(\gt)$ in Eq.
(\ref{eq:orthogonal_decomp}) by replacing $s_j \rightarrow \chi_j$ and
differentiating with respect to $\chi_i$ (and is truncated here at second order
because we assumed, for simplicity, that genetic interactions are limited
to that order).

Converting $\phi$s to $\chi$s using the relation (\ref{eq:chi_to_phi}), we find
$\dot{\chi}_i = (1-\chi_i^2)\dot{\phi}_i$, correct to the leading order. For the
discussion below, it will be useful to derive equations for $\chi_i$ and
$\chi_{ij}$ also to the sub-leading order
\begin{eqnarray}
\label{eq:QLE_cumulants}
\dot{\chi}_i &=
&\sum_j \chi_{ij} \left [ \fitLEcoeff_j- \chi_i \fitcoeff_{ij} \right ] 
+\sscale\mathcal{O}(\sscale^2/r^{2})  \\ \chi_{ij}  & =&
\frac{(1-\chi_i^2)(1-\chi_j^2)f_{ij}}{2\fitLEcoeff_{i}\chi_i +
2\fitLEcoeff_{j}\chi_j + \rec c_{ij}}+\mathcal{O}(\sscale^2/ r^{2}) \  \ \ \ 
\mathrm{for} \ \ i \neq j \nonumber
\end{eqnarray}
In QLE, correlations $\chi_{ij}$ between loci ($i\neq j$) are determined by the
balance between epistatic selection and recombination. (Note, in contrast, the diagonal
elements $\chi_{ii} = \langle \locus_i^2\rangle - \langle \locus_i\rangle^2 =1 -\chi_i^2$ are
determined by the allele frequencies.)

\citet{Wright_Genetics_1931} showed that in {\it linkage equilibrium}, the
dynamics of allele frequencies are driven by the gradient in mean fitness. The
result can be generalized to include correlations between loci arising in QLE.
Starting with the exact equation for the allele frequency dynamics and using our
parameterization of $P(\gt)$ via the ``fields" $\phi_i$ given in
Eq.~(\ref{eq:logP}), we find
\begin{eqnarray}
\label{eq:Wright}
\dot{\chi}_i &=& \langle s_i F \rangle -\chi_i \langle  F \rangle
=\partial_{\phi_i} \langle  F \rangle \ \approx \ \sum_j
\partial_{
\phi_i} \chi_j \partial_{\chi_{j}} \mfit\\ \nonumber
&=& \sum_j
\chi_{ij} \partial_{\chi_j}\mfit 
\end{eqnarray}
where we have used the chain rule of differentiation and the fact that  
${\partial \chi_j \over \partial {\phi_i}} = \chi_{ij}$ following directly from
Eq.~(\ref{eq:gen_of_cumulants}). The correlation matrix $\chi_{ij}$ acts as a 
mobility matrix for allele frequencies. The non-diagonal entries of order
$\sscale/\rec$ imply that selection on locus $j$, via the correlation with
locus $i$, affects the rate of change of $\chi_i$. Eq.~(\ref{eq:Wright})
describes the dynamics of allele frequencies as the population ascends
Wright's ``adaptive landscape". While allele frequencies still evolve to
maximize $\mfit$, their dynamics now are coupled by correlations captured in
the off-diagonal terms of $\chi_{ij}$.

The key point emerging from the
analysis of the weak selection/rapid recombination limit is the remarkable
simplicity of multi-locus dynamics: the $2^L$ ordinary differential equations for
all cumulants or equivalently for all genotypes are reduced to $L$ differential
equations describing the dynamics of allele frequencies. Higher order cumulants
are slaved to allele frequencies and can be obtained by solving algebraic
equations defining the $L$ dimensional quasi-linkage-equilibrium manifold.  The
distribution of genotypes in the population can therefore be parameterized by
time-dependent allele frequencies, with all other features of the distribution
constrained by the QLE equations. In mathematical terms, the dynamics of genotype
distribution are approximately reducible to the dynamics on the ``center
manifold" formed by the set of allele frequencies \citep{Guckenheimer:1997p36896}. Within
the QLE approximation, population averages of any trait $G(\gt)$ can be
parameterized by $\{\chi_1(t), \ldots, \chi_L(t)\}$ and the time-derivative of 
the trait mean is therefore given by
\begin{equation}
\label{eq:QLE_trait_mean_dyn}
\begin{split}
\frac{d}{dt}\langle  G \rangle
&\approx \sum_i \partial_{\chi_i} \langle  G \rangle \partial_t\chi_i(t)=
\sum_{ij}
\chi_{ij} \partial_{\chi_i} \langle  G \rangle \partial_{\chi_j} \langle  F \rangle -2\mu\sum_i \chi_i\partial_{\chi_i}
\langle  G \rangle
\end{split}
\end{equation}
where we have restored the contribution of mutations through its effect on allele 
frequencies as it appeared in Eq.~(\ref{eq:general_allele_freq_dyn}).
This result has a very simple interpretation: The rate of change of the trait
mean is the product of the rate of change of allele frequencies through
selection and the susceptibility of the trait mean to the allele frequency. The
second term accounts for the effect of mutation on the trait mean. Since the
first term is the additive covariance between fitness and the trait $G$,
this equation is the analog of Eq.~(\ref{eq:price_equation}) in a recombining
population. The QLE approximation breaks down when recombination is not
sufficiently rapid to confine the genotype distribution to the $L$ dimensional
manifold defined by quasi-steady correlations between loci. This breakdown will
be discussed in more detail below.

\subsection{Additive genetic variance and Fisher's theorem in QLE}
Fisher's theorem in sexual populations posits that the rate of mean fitness
increase is equal to the additive variance. We will now discuss how Fisher's
theorem emerges from Eq.~(\ref{eq:general_mean_fitness_dyn}) and how it compares
with Eq.~(\ref{eq:QLE_trait_mean_dyn}) which obviously can be used to calculate
$d\mfit/dt$. Additive variance is typically defined as the variance captured by
a linear model of the form
\begin{equation}
\label{eq:additive_model}
\fit_A(\gt) = \acoeff_0 + \sum_i \acoeff_i\locus_i
\end{equation}
where the coefficients are determined by minimizing
\begin{equation}
\label{eq:sigma_I}
\sigI=\sum_\gt (\fit_A(\gt)-\fit(\gt))^2 P(\gt, t) \ .
\end{equation}
The remaining variance $\sigI$ is commonly called \emph{epistatic} or
\emph{interaction} variance. Minimization yields $\acoeff_0 = \mfit - \sum_i
\acoeff_i\chi_i$ with $\acoeff_i$ determined by the linear equation
\begin{equation}
\sum_j \chi_{ij}\acoeff_{j}
= \langle \locus_i F  \rangle -\chi_i  \mfit=\partial_{\phi_i} \mfit
\end{equation}
We have seen the right hand side of this equation already in
Eq.~(\ref{eq:Wright}): it is the contribution of selection  to $\dot{\chi}_i$.
In the high recombination limit, $\partial_{\phi_i} \mfit
\approx \sum_{j} \chi_{ij} \partial_{\chi_j}
\mfit$. Hence the additive fitness coefficients (defined by linear regression)
are $\acoeff_i =   \partial_{\chi_i} \langle  \fit
\rangle$, which is accurate to order $\sscale/\rec$.
The additive variance therefore is 
\begin{eqnarray}
\sigA &=& \sum_{ij}\acoeff_i\chi_{ij}\acoeff_j \approx \sum_{ij} 
\chi_{ij} \partial_{\chi_i} \mfit \partial_{\chi_j} \mfit+\sscale^2\mathcal{O}(\sscale^2/r^{2})
\end{eqnarray}

Recalling the QLE equation for mean trait dynamics, Eq.
(\ref{eq:QLE_trait_mean_dyn}), and using fitness as a trait, we have
\begin{equation}
\label{eq:QLE_fit_mean_dyn}
\begin{split}
\frac{d}{dt} \langle  F \rangle
&\approx \sum_{ij}
\chi_{ij} \partial_{\chi_i} \mfit \partial_{\chi_j} \mfit
-2\mu\sum_i \chi_i\partial_{\chi_i} \mfit
\end{split}
\end{equation}
and comparing to the definition of $\sigma_A^2$ we arrive at the generalized Fisher's ``Fundamental Theorem"
\begin{equation}
\label{eq:FT_centermanifold}
\frac{d}{dt} \mfit  = \sigA - \mu \Delta_\mu +\mathcal{O}(\bar{\sigma}^4/\rec^2)
\end{equation}
which limits growth of fitness to the \emph{additive} variance. 
Comparing to the general expression for mean fitness given before in Eq.
(\ref{eq:general_mean_fitness_dyn}) we see that the loss in fitness due to
disruption of favorable combinations of alleles though recombination
exactly cancels the epistatic $\sigma_I^2 = \sigma^2 - \sigma_A^2$  part of total variance. 
In other words, in a sexually reproducing species the uncertainty
in the phenotype of the offspring in relation to that of its parents limits the effect
of selection to the additive component of variance. The latter is that
genetic component of the trait that ``survives'' reshuffling of genes by
reassortment and recombination which depends on the genetic distance to the
mate. Hence, this decomposition of genetic variation in additive and
non-additive components is explicitly population dependent.

One must of course remember that the generalized Fisher's law as stated only
holds in this rapid recombination/weak selection limit and only after
correlations have relaxed to their steady QLE values. During the initial
transient towards QLE or at low recombination rates mean fitness can exhibit very
different dynamics. The meaning of Fisher's theorem has been subject to extensive
discussion in the literature
\citep{Feldman:1970p33774,Ewens:1989p11361,Price:1972p11365,Frank_1992,Edwards:1994p11406}
caused by Fisher's insistence that {\it his statement} was exact.
\citet{Price:1972p11365} in particular suggested that Fisher's intention was to
describe not the total rate of change of mean fitness, but only the ``partial
rate" due to change in allele frequencies: i.e. just 1st term on the r.h.s.~of
Eq.~(\ref{eq:QLE_trait_mean_dyn}). The ``theorem" would in that case become an
exact statement, but not a very useful one! Following \citet{Kimura:1958} and
\citet{NAGYLAKI:1993p2698} our Eq.~(\ref{eq:FT_centermanifold}) sticks to  $d
\mfit/dt$ so that the generalized Fisher's theorem is an unambiguous, but
approximate statement. The above analysis assumed that the population is subject
to a constant fitness function and the mean fitness provides a useful
measure of adaptation. If the fitness function itself depends on time, the
increase in mean fitness due to adaptation of the population is superimposed with
the dynamics of the fitness function. In the latter case, an unambiguous measure
of adaptation, the fitness flux, can be defined in analogy to fluctuation
theorems of non-equilibrium statistical mechanics \citep{Mustonen:2010p36757}.

The off-diagonal terms in the additive variance $\acoeff_i\chi_{ij} \acoeff_j$
have interesting implications for the evolution of recombination: if two alleles
that are selected with the same sign ($\acoeff_i\acoeff_j>0$) are anti-correlated
($\chi_{ij}<0$), the rate of adaptation is smaller than it would be in linkage
equilibrium. This is the basis for the often made statement that recombination
accelerates adaptation by reducing negative linkage disequilibria and thereby
increasing the additive variance \citep{Barton:2005p982}. There is, however, an
additional effect of recombination on adaptation that is not captured by
deterministic multilocus dynamics and  is likely to be more important:
Recombination greatly increases the likelihood that a novel beneficial mutation
establishes and ultimately fixates in the population
\cite{Barton:1995p3540,Fisher_1930,Muller_AmericanNaturalist_1932,Neher:2010p30641}.
Thereby the number of simultaneously polymorphic loci is increased, which in turn
increases the fitness variance and speeds up adaptation. The reason for this is
again that recombination breaks down negative linkage disequilibria (a tendency
of beneficial alleles to be anti-correlated), which are generated by chance and
amplified by selection \citep{Barton:2005p982}. Analysis of this phenomenon
requires going beyond QLE (see below).

\section{Finite population drift and Wright's mutation/selection/drift equilibrium.}
\label{sec:drift}
So far our formulation of the genotype dynamics Eq.~(\ref{eq:genotype_dynamics})
and the dynamics of allele frequencies Eq.~(\ref{eq:QLE_cumulants}) was
deterministic, i.e.~we neglected random drift. Random drift is a consequence of
the stochastic nature of birth and death in a finite population of size $N$.
In the simplest models of stochastic population genetics -- called
Fisher-Wright models -- stochasticity is introduced by resampling the
population from a multinomial distribution parameterized with the current
genotype (or gamete) frequencies each generation. 

We have seen above that the
genotype frequency distribution can be parameterized by allele frequencies when
recombination is rapid and we shall discuss now how resampling of genotypes
leads to stochastic contributions to the dynamics of allele frequencies and
cumulants. For alleles that are present in large numbers, the relative sizes of
fluctuations due to resampling are small and random drift can be accurately
described by a diffusion approximation \citep{Kimura:1964p3388}.
To derive a diffusion equation for allele frequencies, we generalize the ordinary differential
equations Eq.~(\ref{eq:QLE_cumulants}) to stochastic differential equations
(Langevin equations \cite{WGardiner:2004p36981}). For a finite time step $\Delta
t$, one has
\begin{eqnarray}
\label{eq:langevin}
\chi_i(t+\Delta t) &= &\chi_i(t)+\Delta t\left[\sum_j \chi_{ij}\partial_{\chi_j}
\mfit -2  \mu \chi_i\right] + \sqrt{\Delta t} \zeta_i(t) \\
\chi_{ij}(t+\Delta t) &=& \chi_{ij}(t) + \Delta t \left[
(1-\chi_i^2)(1-\chi_j^2)\fitcoeff_{ij}-\rec\xo_{ij} \right] + \sqrt{\Delta t}
\zeta_{ij}(t)
\end{eqnarray}
where we have neglected terms much smaller than $\rec\xo_{ij}$ in
the relaxation rate of $\chi_{ij}$. $\zeta_i(t)$ and $\zeta_{ij}(t)$ are 
white noise terms with zero mean and a covariance matrix determined by the
multinomial sampling of the genotypes. One finds
\begin{eqnarray}
\langle \zeta_i(t) \zeta_j(t')
\rangle &=& \frac{\chi_{ij}}{N}\delta(t-t')\\
\langle \zeta_{ij}(t) \zeta_{ij}(t')
\rangle &\approx& \frac{(1-\chi_i^2)(1-\chi_j^2)}{N}\delta(t-t')
\end{eqnarray}
while other covariances are of order $\sscale/\rec$ or smaller, see
Appendix \ref{sec:app_wright}. The joint stochastic dynamics of allele
frequencies and the correlation between loci has been studied by
\citet{Ohta:1969p4413} using a two-locus model. Here, we study a multi-locus
model making the simplifying assumption that the recombination is faster than all other
processes.

In this case, the 2nd order cumulants relax much faster than allele frequencies
change and we can solve the equation for $\chi_{ij}$ assuming fixed $\chi_i$.
The solution can be decomposed into a deterministic component due to the competition between
epistatic selection and recombination and a stochastic component.
\begin{equation}
\chi_{ij}(t) = \frac{\fitcoeff_{ij}(1-\chi_i^2)(1-\chi_j^2)}{\rec\xo_{ij}} +
\delta \chi_{ij}
\end{equation}
The deterministic component is the familiar QLE value from Eq.~(\ref{eq:QLE_cumulants}),
while the stochastic component $\delta \chi_{ij}$ has an auto-correlation
$\langle \delta\chi_{ij}(t)\delta\chi_{ij}(t+\Delta t)\rangle =
\frac{(1-\chi_i^2)(1-\chi_j^2)}{2N\rec}e^{-r\Delta t}$, see Appendix
\ref{sec:app_wright}. We will now use this result to study Langevin equation
for $\chi_i$. We have to distinguish the case where the deterministic
component to $\chi_{ij}$ dominates over the stochastic term or vice versa. 
In order to compare the stochastic to the
deterministic term, we have to average the former over the time scale of the
dynamics of $\chi_i$ given by the inverse of $\frac{\partial
\mfit}{\partial \chi_i}\approx \fitLEcoeff_i$. Recalling that the equilibrium
value of the $\chi_i \approx 1-\mut/\fitLEcoeff_i$, we find that the
deterministic contribution to $\chi_{ij}$ dominates if $N\mut\gg 1$
and $\fitcoeff_{ij}\gg \mut$.
In the opposite limit, the stochastic contribution $\delta\chi_{ij}$ will
affect the dynamics of $\chi_i$ more strongly than the deterministic one. 
We will now show how the equilibrium distribution of allele frequencies is
affected by correlation between loci in these two cases.

\subsection{Wright's equilibrium in the QLE approximation}
Assuming we can neglect the stochastic contribution to $\chi_{ij}$, the
Langevin equation for the $\chi_i$ (interpreted in the \^Ito sense)
corresponds to the following forward Kolmogorov equation for the dynamics
of the probability distribution of allele frequencies by $Q(\{ \chi_i\}, t)$ 
\citep{WGardiner:2004p36981}
\begin{equation}
\label{eq:FPE2}
\partial_t Q(\{\chi_i\},t) = \sum_i \partial_{\chi_i} \left[
\frac{1}{2N}\sum_{j} \partial_{\chi_j} (\chi_{ij}Q(\{\chi_i\},t) ) +
Q(\{\chi_i\},t) \left ( 2\mu\chi_i-\sum_j \chi_{ij}\partial_{\chi_j} \mfit \right )\right]
\end{equation}
This multilocus version of the diffusion equation for allele
frequencies in linkage equilibrium (no correlations) appears already in
\citet{Kimura:1955p36295}. It has a steady solution where all probability flux vanishes,
i.e.~where the term in brackets is zero for each $i$. In complete linkage equilibrium, 
the matrix $\chi_{ij}$ is diagonal and different allele
frequencies decouple. One obtains the equilibrium distribution 
\begin{equation}
\label{eq:wright equilibrium}
  Q (\{ \chi_i \} )= C e^{  2N \fit(\{\chi_i\})} \prod_i (1- \chi_{i}^2
  )^{2N\mu-1}\ ,
\end{equation}
where $\fit(\{\chi_i\})$ is the mean fitness evaluated in linkage equilibrium
obtained by replacing each $\locus_i$ by its moment $\chi_i$ in
Eq.~(\ref{eq:orthogonal_decomp}). The term $e^{ 2N\fit(\{\chi_i\})}$ is
analogous to the contribution of energy to a Gibbs measure, while $\prod_i (1-
\chi_{i}^2 )^{2N\mu-1}$ plays the role of an entropy. Note that for $2N\mu<1$, the
distribution is singular at $|\chi_i|=1$. In the opposite case $2N\mu>1$, $Q
(\{ \chi_i \} )$ vanishes if any of the $|\chi_i|=1$. Instead $Q(\{ \chi_i \} )$ has
a maximum in the interior of the hypercube defined by $|\chi_i|< 1$. 

The corresponding solution for QLE, where
$\chi_{ij}$ has small but steady off-diagonal entries is derived in
Appendix \ref{sec:app_wright} with the result:
\begin{equation}
\label{eq:QLE_wright equilibrium}
Q(\{\chi_i\}) = C e^{2N\mfit + 4N\mu\sum_{i<j}
\frac{\fitcoeff_{ij}\chi_i\chi_j}{\rec\xo_{ij}}}
\prod_{i=1}^L(1-\chi_i^2)^{2N\mu-1}
\end{equation}
The genotype distribution assumes this exponential (Boltzmann) form
$\sim e^{N\mfit}$ since the mobility matrix $\chi_{ij}$ is proportional to the
auto-correlation of the genetic drift. Eq.~(\ref{eq:QLE_wright equilibrium})
provides a systematic extension of Wright's equilibrium to QLE, which appears
to be a new result.

\subsection{Wright's equilibrium with stochastic linkage disequilibrium}
In absence of epistasis or in cases where selection is weak or comparable to
the strength of genetic drift (diffusion constant), the deterministic
expectation for $\chi_{ij}$ is small compared to its fluctuations. The coupling
between different allele frequencies in Eq.~(\ref{eq:langevin}) has therefore
fluctuating sign and acts as an additional noise source with auto-correlation
time $(\rec\xo_{ij})^{-1}$. Such an increased noise level increases the
diffusion constant in the Fokker-Planck equation for each of the $\chi_i$ by a
factor
\begin{equation}
\label{eq:Ne}
\frac{N}{N_e} = 1+\frac{1}{2}\sum_{i\neq
j}(1-\chi_j^2)\left(\frac{1}{\rec\xo_{ij}}\frac{\partial \mfit}{\partial
\chi_j}\right)^2
\end{equation}
This increase in diffusion constant is often phrased as a reduction in
effective population size $N_e$ and is known as a manifestation of the
Hill-Robertson effect \citep{Hill:1966p21029}. Note that the correction has the structure of
the additive variance in fitness where each term is compared to the square of
the recombination rate between the loci $i$ and $j$. This result was derived in
the context of fixation probabilities of novel mutations in
\citet{Barton:1995p3540}. It has been shown that this effective increase in the
diffusion constant through stochastic correlations of loci can select for
increased recombination rates \citep{Barton:2005p982}.

\subsection{Equilibration towards a steady state}
The approach to the equilibrium distribution is governed by the
smallest non-zero eigenvalue of Eq.~(\ref{eq:FPE2}). For $2N\mu >1$ and smooth
fitness landscapes, this relaxation rate is governed by the larger of 
$\mu$ and the scale of selection on individual loci. The corresponding time scales can
be very long.
Furthermore, if different parts of sequence space are separated by
fitness valleys (``energy barriers"), relaxation to the steady state can take
exponentially long \citep{Weissman:2010p37077}.

Similar equations for the distribution of allele frequencies apply in the
context of spatially structured populations, in which case the role of
mutation is played by migration of individuals. The latter problem was the
subject of work by \citet{Wright:1932p40300}. Migration rates and the
associated influx of foreign alleles are often much larger than mutation rates
and rapid equilibration is plausible.

\section{Breakdown of QLE}
\label{sec:QLE_breakdown}
QLE greatly simplifies the dynamics of the genotype distribution, but the
perturbation theory nature leaves one with the question about the range of its
validity. In particular, we know from Statistical Physics that Gibbs measures of
the form of Eq.~(\ref{eq:logP}) can lead to a so-called glass transition where
the structure of the distribution changes qualitatively. Below the glass
transition, different realizations of the system have a non-vanishing probability
to be (largely) identical, which is quantified by the overlap distribution
(Parisi order parameter, \citet{Mezard:2009p28797}). A related transition in
which the population condenses into a small number of genotypes  is driven 
by the competition between epistasis and recombination. It occurs already
in the deterministic mean field setting and is discussed
below in Section \ref{sec:QLE_breakdown_deterministic}. QLE can also become 
unstable at low recombination rates even in the absence of epistasis because of 
the discreteness of  contributions of individual loci in a finite genome. The instability in that case is driven by 
fluctuations due to finite population size and is discussed
in Section \ref{sec:QLE_breakdown_stochastic}.

\subsection{Infinite $N$ and $L$ limit: Alleles vs.~Genotypes}
\label{sec:QLE_breakdown_deterministic}
To gain some heuristic insight into the range of validity of the perturbation
expansion in $\sscale/r$, it is useful to study the following coarse-grained
``quantitative genetic'' version of QLE which yields an explicit criterion for 
the validity of QLE \citep{Neher:2009p22302}. Instead
of following the entire genotype distribution, consider the joint distribution
$P(A,E,t)$ of additive $A$ and epistatic $E$ contributions to fitness defined via
$A=F_A(\gt)$ (comp.~Eq.~(\ref{eq:additive_model}))  and $E=F-A$. Hence additive
and epistatic contributions are defined with reference to the current distribution of
genotypes. The joint distribution of $A$ and $E$ evolves according to
\begin{equation}
\label{eq:MFT}
\partial_t P(A,E,t) = (A+E-\langle A \rangle-\langle E \rangle)P(A,E,t) +
\rec\left(\rho(E)\int dE'\, P(A,E',t)-P(A,E,t)\right)
\,
\end{equation}
where $\langle A \rangle$ and $\langle E \rangle$ are the mean additive and
epistatic fitness in the population. Here we have assumed that the epistatic
fitness of novel recombinants is independent of its parents and given by a
random sample from the density of possible epistatic fitness values $\rho(E)$
(the ``house-of-cards" model, \citep{Kingman:1978p31199}). We assume $\rho(E)$ to be a Gaussian with
the variance equal to $\sigma_I^2 $ - the epistatic component of fitness
variance defined in Eq.~(\ref{eq:sigma_I}). Additive fitness of
recombinants is a random sample from the current distribution of additive 
fitness in the population, i.e.~the marginal $\int
dE'\, P(A,E',t)$. The model does not include finite population size effects and
assumes that both $A$ and $E$ are from a continuous distribution. The latter
implies that the number of loci $L$ that contribute to fitness is very large,
while the individual contributions of loci are small (comp.~Section
\ref{sec:geno_pheno}). In this sense, it is a deterministic mean-field model.

Equation (\ref{eq:MFT}) has a factorized solution
$P(A,E,t)=\theta(A,t)\omega(E)$ with
\begin{equation}
\label{eq:QGqle}
\theta(A,t) = \frac{1}{\sqrt{2\pi
\sigA}}e^{-\frac{(A-\langle A \rangle)^2}{2\sigA}}\quad\mathrm{where}\quad
\frac{d}{dt}\langle A \rangle =\sigA
\end{equation}
\begin{equation}
\nonumber
\omega(E)  =\frac{\rec\rho(E)}{\rec+\langle E \rangle-E} \quad\mathrm{where}\quad \langle E \rangle
= \int dE\,E\omega(E)
\end{equation}
The mean epistatic fitness is determined by enforcing the normalization of $\omega(E)$,
i.e.~$\int dE\, \omega(E) =1$. Note that this solution is a QLE solution: Fitness
increases with a rate given by the additive variance, while the epistatic
contribution to fitness is steady with a magnitude controlled by recombination.
Unlike Eq.~(\ref{eq:QLE_cumulants}), Eq.~(\ref{eq:QGqle}) implies a condition on
$\rec$ and the density of states: $\rho(E)$ has to vanish for $E\geq
\rec+\langle E \rangle$. Otherwise, $\omega$ is not normalizable. The density of
states $\rho(E)$ is typically of Gaussian form, and given $2^L$ states has its maximum
at $E_{max}\approx \sigI
\sqrt{2L\log 2}$ (if $N\ll 2^L$, as will be generically the case, $E_{max}
\approx \sigI \sqrt{2\log N}$). Hence QLE is expected to break down at $\rec_c
\approx E_{max}-\langle E \rangle\approx E_{max}$.

\begin{figure}[tp]
\begin{center}
  \includegraphics[width=\columnwidth]{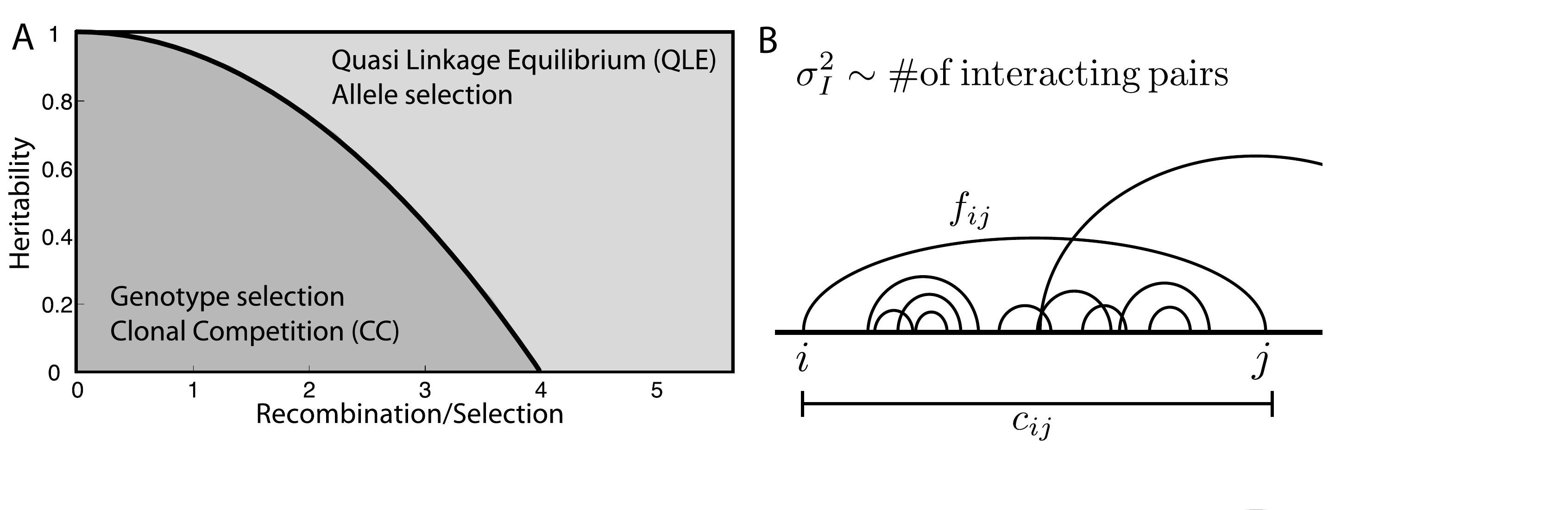}
  \caption[]{Genetic interactions and the breakdown of QLE. Panel A shows the range of validity
  of QLE as a function of $\bar{\sigma}/\rec$ and the heritability, i.e. the
  ratio of additive variance to the total fitness variance. Below the transition line, strong linkage equilibrium is expected and selection operates on genotypes rather than alleles. 
  Panel B shows a illustration of a possible ``interaction graph" of polymorphisms on a block of chromosome. Loci interact
  with nearby loci, as well as with distant loci outside the block. The epistatic fitness variance solely within the block, 
  i.e.~averaged over the rest of chromosome, is proportional to the number of interaction terms (arcs in the figure). }
  \label{fig:phasediagram}
\end{center}
\end{figure}

The dynamics of the distribution of $P(A,E)$ change dramatically as $\rec$ falls
below $\rec_c$: For $\rec>\rec_c$ no genotypes with $E\geq \rec+\langle E
\rangle$ exist. Hence all genotypes are destroyed by recombination and
short-lived. At $\rec<\rec_c$, however, many genotypes with $E\geq \rec+\langle E \rangle$
exist which can outrun recombination and grow exponentially. The genotype distribution
is no longer a product of additive and epistatic parts, but contains clones which
are populated by many individuals. Selection now operates on the entire genotype
over many generations and the relevant dynamical quantities are now clone sizes rather
than allele frequencies, which are slaved to the performance of the clones. The
alleles that make up the most successful genotype will fixate, not necessarily
those with the most favorable additive effect. The transition line between the
two regimes is sketched in Fig.~\ref{fig:phasediagram} with the ratio of
recombination to selection on the x-axis and the ``heritability" - the ratio of
the additive variance to the total variance  $h^2 = \sigma_A^2 /(\sigma_A^2
+\sigma_I^2) $ - on the y-axis. (Note that heritability also measures the
correlation between fitness of a recombinant offspring and parental mean
\citet{Lynch:1998p8721}.) At low recombination rates and strong epistatic
interactions, selection operates on genotypes while at high recombination rates or in absence
of epistasis, selection operates on the additive effects of alleles. The
distinction between genotype and allele selection regimes goes back to
\citet{Franklin:1970p13820,Slatkin:1972p13822}, who showed that a related
transition occurs in models with strong heterozygote advantage. The regimes of
allele and genotype selection are summarized in Fig.~\ref{fig:phasediagram}. In
absence of epistatic interactions or heterozygote advantage, a similar
condensation phenomenon occurs only at very low outcrossing rates
$\rec=\mathcal{O}(N^{-1})$ \citep{Rouzine:2005p17398}.

The condensation of genotypes goes along with a dramatic speed up of the
allele frequency dynamics: In QLE (allele selection) each allele frequency is driven by
an effective additive coefficient $\acoeff_i \sim \sscale/\sqrt{L}$
(each $\acoeff_i$ accounts for $\sim L^{-1}$ of the additive variance
$\sigA$). When selection operates on genotypes, the time scale of selection
is driven by fitness differences between individuals, which are of order
$\sscale$. The rate of change of allele frequencies is therefore greater by a
factor $\sqrt{L}$ which could be a large effect \citep{Neher:2009p22302}. We emphasize that the stationarity
of the distribution of the epistatic component of fitness $\omega (E)$ holds only on time scales
short compared to that of the allele frequency dynamics. 

The simple picture of the transition in a facultatively outcrossing species can
also apply to blocks of chromosome in obligate sexuals. Consider a block
that harbors $l$ loci spread over a map distance $\xo$ (on average $\xo$ recombination
events within the block per generation). If epistatic fitness within the block
exceeds $\xo$, QLE will break down since individual haplotypes will be amplified
by selection above less fit recombinants. Whether such local breakdown of QLE
will occur depends on how fitness variance and recombination rate depend on the
block size. The recombination rate is proportional to the block size, and,
assuming constant density of polymorphism, will be proportional to the number
$l$ of polymorphic loci. Similarly, the additive variance is proportional to $l$
and the r-m-s therefore $\sim \sqrt{ l}$. The epistatic variance within the block scales with the number of interactions
between loci within the block, as illustrated Fig.~\ref{fig:phasediagram}B. Any
given locus will interact only with a fraction of all other loci,
i.e.~$\fitcoeff_{ij}$ is sparse, and the number of interactions between loci
within a block depends on whether these sparse interactions tend to be local or not. If any two loci are
equally likely to interact, the number of interactions within the block is $\sim
l^2$, so that r-m-s epistatic fitness is $\sim l \sim \xo$. Hence the ratio of
recombination within the block and the epistatic fitness are independent of the
block size and QLE is either globally stable or unstable. A different conclusion
is reached if interactions are local and each locus interacts with $k$ nearby
other loci. As before additive fitness $\sim \sqrt{l}$, but the number of
interactions within the block is $\sim lk$. Hence the typical epistatic fitness
is $\sim \sqrt{lk}$, which decreases less fast than $\xo$ as the block length is
decreased. We therefore expect that QLE is unstable on scales below a critical
block size $l_c$, where local epistasis overwhelms rare recombination. This local
selection on coadapted haplotypes can coexist with establishment of QLE on longer
genomic scales \citep{Neher:2009p22302}. We will come back to the recombination
and selection on different chromosomal scales in the Discussion.

\subsection{Validity of QLE for finite $N$ and $L$.}
\label{sec:QLE_breakdown_stochastic}
The above discussion of the breakdown of QLE focussed on the competition between
genetic interactions driving and recombination destroying correlations in the
limit where fluctuations are negligible and contributions of individual loci are
small. We will now discuss how the discrete contributions of individual loci and
the number fluctuations in finite populations can drive populations off the QLE
manifold. This problem has a long history in population genetics and was mainly
discussed for scenarios without genetic interactions, i.e.~on the line where
the heritability equals 1 in Fig.~\ref{fig:phasediagram}A. In this case the only
source of correlations are the initial condition or fluctuations. \citet{SMITH:1968p6851} showed that without
genetic interactions and with no correlations in the initial condition,
correlations do not develop in an infinite population at any recombination
rate, in accordance with Fig.~\ref{fig:phasediagram}A. However, novel mutations
arise in single copies on random genomes, giving rise to correlations: QLE has to be stable with 
respect to these perturbations.

The hallmark of the QLE approximation are slowly changing allele frequencies
and steady and perturbative correlations between loci. The latter will only be true, 
if the correlations relax, i.e.~are governed by an equation of the form
$\dot{\chi}_{ij} = \beta - \alpha\chi_{ij}$ with $\alpha =
2(\fitLEcoeff_i\chi_i+\fitLEcoeff_j\chi_j) + \rec \xo_{ij}>0$ (ignoring
mutations). Hence the QLE state
is unstable if $-2(\fitLEcoeff_i\chi_i+\fitLEcoeff_j\chi_j)>\rec \xo_{ij}$. 
In that case any small deviation from $\chi_{ij}=0$, which could be due to stochastic
fluctuations, will grow. This effect has important implications for the evolution
of recombination: Consider two closely linked loci at which beneficial mutations
happen. Both novel mutation exists initially as a single copy ($\chi_i\approx
-1$) and will most likely reside in different individuals, i.e.~are
anti-correlated or in negative equilibrium. Selection will now amplify the
initial $\chi_{ij}$ if $2\fitLEcoeff_i+2\fitLEcoeff_j>\rec\xo_{ij}$, generating
predominantly negative LD. This growth of correlations due to selection on
individual loci slows down adaptation and can result in the loss of beneficial
alleles. This phenomenon is known as Hill-Robertson interference and it
contributes to potential benefits of sexual reproduction
\citep{Hill:1966p21029,Barton:1995p3540,Barton:2005p982}. While this
cumulant based approach to interference between sweeping loci is tractable
for few loci, it becomes intractable in populations in which many sweeping loci
are tightly linked \citep{Cohen:2005p5007,Rouzine:2005p17398,Neher:2010p30641}.

\subsection{Cumulant analysis beyond QLE} 
Even though the QLE approximation breaks down when correlations are no longer
slaved variables, the cumulant expansion can be useful to study the short term
dynamics of systems with a small number of loci, in particular if the initial
conditions are such that higher order cumulants are small. Furthermore, if
only a few isolated pairs of tightly linked loci are present, cumulants between
these pairs can be treated as dynamical variables, while all other pairs
for which the $\chi_{ij}$ are stable are treated in QLE. Such an analysis has
for example been performed by \citet{Stephan:2006p12174} to study LD patterns
between neutral markers following a selective sweep.

Explicit modeling of stochastic multi-locus systems typically requires computer
simulations, which are computationally expensive when the number of loci or the
population size is large. However, making use of the Fast-Fourier Transformation
on the $2^L$ dimensional genotype space, one can speed up such simulation from a
runtime that scales as $8^L$ to $3^L$. The FFT allows to calculate and reuse the
frequency of subsets of loci from which the distribution of recombinant genomes
can be assembled. An efficient implementation of multi-locus evolution for
arbitrary fitness functions and genetic maps is available from the author's
website. Cumulant equations to higher order involve ``book-keeping'' of many
terms and is best done with computer algebra systems. A package for
Mathematica\textregistered~ has been developed by \citet{Kirkpatrick:2002p2655}.
A implementation for Maple\textregistered~ is available from the authors.

\section{Discussion}
We have presented a review of the dynamics
of multi-locus genotype distributions and the resulting dynamics of quantitative
traits. We focused in particular on how the distribution of genotypes can be
parameterized by allele frequencies in the weak selection/fast recombination
limit. This description extends beyond ``beanbag genetics" allowing also for weak
correlation (i.e.~linkage disequilibrium) between loci. The central element is
the Quasi-Linkage Equilibrium approximation pioneered by Kimura. QLE emerges as a
perturbation expansion in the weak selection/rapid recombination limit similar to
high-temperature expansion in statistical physics. In a suitably defined system,
the population genetics can be classified by the ratio of the strength of
selection and the rate of recombination and the degree to which the fitness
variation is additive or epistatic, see Fig.~\ref{fig:phasediagram}. At high
recombination and additivity, QLE is an accurate approximation. This regime is
separated from a regime at low recombination and strong epistasis, where QLE
breaks down and the population condenses into a few fit genotypes.

Our exposition assumes a panmictic, random mating, and haploid population. While
the former are pretty common assumptions, assuming haploidy in recombining
population might raise objections. Our aim was to discuss the interplay between
selection, genetic interactions and recombination in multi-locus systems.
Dominance is a special kind of genetic interaction, where a locus interacts with
itself, giving rise to additional non-linearities. These non-linearities can
stabilize loci at intermediate allele frequencies, a process not possible in
haploid populations. The effects of dominance, however, are well understood at
the single locus level, as well as when many loci with heterozygote advantage are
close to each other \citep{Franklin:1970p13820}. Within QLE, the dynamics of
allele frequencies in diploid populations is still relaxational and maximizes the
mean diploid fitness. A full parametrization of the diploid populations and
diploid fitness requires a straightforward, if somewhat tedious, generalization:
To represent diploid one should (i) double the number of loci, (ii) define a
genetic ``transfer function", $C(\{\xi\})$, that represents meiotic crossover of
the parental genomes, (iii) extend the fitness function $F(\gt)$ to $2L$
hypercube to parameterize  the $3^L$ states (homo- and heterozygocity at $L$
loci). Another simplification of our exposition was the
use of the continuous time description, in contrast to the more common discrete
generation formulations of population genetics. Continuous time formulations
assume that the population changes little in one generation. If this is the case,
the results are completely equivalent and one can make use of calculus instead of
recursions and difference equations.

The QLE approximation will often be appropriate for panmictic populations where
genetic variation is replenished by de novo mutations. In this scenario, novel
mutations establish if they blend in well with genetic make up of the population.
This is manifest in the QLE equation (Eq.~(\ref{eq:QLE_cumulants})) were
alleles are selected on the basis of their additive effect, i.e.~their effect on
fitness marginalized over the distribution alleles at other loci in the population. The fitness of individual
genotypes is not relevant to the evolutionary dynamics, since genotype
frequencies are determined by allele frequencies (and sampling noise in finite
populations). This issue was recently discussed in
\citet{Livnat:2008p20709}.

A very different evolutionary dynamics follows a hybridization event
\citep{Nolte:2010p32541,Barton:2001p22181,Orr:1995p4894}, i.e.~a situation when
two strains of one species that have been evolving in isolation for some time
come in contact again. The two strains will differ at many loci and these
differences have never been tested for compatibility. Crossing two such diverged
strains can result in a phenotypically diverse population from which novel hybrid
species can emerge \citep{Nolte:2010p32541}. Such speciation after hybridization
is similar to the clonal population structure observed in theoretical models of
the selection dynamics after hybridization \citep{Neher:2009p22302}. In this
regime of clonal competition, the allele frequencies are slaved to the dynamics
of the clones and the average effect of an individual mutation affects the fate
of a clone only very mildly. The lucky accident that produced through
recombination a very fit genotype that contains the allele determines whether the allele can fixate or
not. The possibility of a sharp transition between mixing and not-mixing of two
populations in a hybrid zone has already been described by
\citet{Barton:1983p34506}, who used a model of hybrid-inferiority. In the limit
of large number of contributing loci, there exists a critical ratio of
recombination rate and selection against hybrids, which separates the regimes of
mixing and non-mixing. Similar transitions are expected if the reason for
out-breeding depression is epistasis rather than dominance.

The qualitative differences between the genotype and allele selection regimes
also sheds light on the importance of stochasticity (genetic drift). Allele 
frequencies are
well sampled by $\mathcal{O}(N)$ copies, unless the allele is very young (or
about to go extinct). Stochasticity therefore matters only during the
establishment phase of the allele. As soon as the frequency exceeds
$(N\fitLEcoeff_i)^{-1}\approx \sqrt{L}/N\sscale$, selection dominates. In the
genotype selection phase, however, the founding of each genotype can, if it is
exceptionally fit, change the fate of the population dramatically.

The transition to genotype selection driven by epistatic interaction
is related to spin-glass transition in models for disordered physical systems and
magnets. Within these models, the probability of finding the system in a
particular state $\{\locus_i\}$ is given by
\begin{equation}
\label{eq:spin_hamiltonian}
P(\{\locus_i\}) \sim e^{-\mathcal{H}(\{\locus_i\})/kT} = e^{-{\frac{1}{kT}} \left [{\sum
h_i\locus_i +
\sum_{ij}J_{ij}\locus_i\locus_j+ \ldots} \right ]} \ ,
\end{equation}
and hence completely analogous to Eq.~(\ref{eq:logP}). Such system generically
reside in one of three states: \emph{paramagnetic}, \emph{ferromagnetic},
\emph{glassy}. In the paramagnetic state at high temperature different
parts of the system are uncorrelated, which is analogous to QLE. The
perturbation expansion in $\sscale/\rec$ is very similar to a high
temperature expansion in statistical physics. At low temperature, the behavior
depends on the structure of the Hamiltonian $\mathcal{H}(\{\locus_i\})$. If most
of the $J_{ij}$ have the same sign, the system will go to an energetically
favored ordered state where spins are aligned, giving rise to a ferromagnet. In
this case $\mathcal{H}(\{\locus_i\})$ has one heavily preferred energy minimum,
corresponding to a very fit genotype.

A different low temperature behavior is found when the $J_{ij}$ have erratic
sign. In that case, not all interactions can be in their favorable state
simultaneously and the resulting landscape has many minima and maxima. At low temperature, the
system condenses into one of the minima. This ``spin-glass" phase is
characterized by a non-trivial overlap distribution: Different realizations of
the system, drawn from the ensemble defined by Eq.~(\ref{eq:spin_hamiltonian}),
will fall into clusters of different degrees of similarity (measured by
Hamming Distance). The clusters themselves have subclusters, giving rise to a
hierarchical ultrametric structure \citep{Mezard:1987p21192}. This is in contrast
to the high temperature phase, where most systems share a typical number of
sites. These qualitatively different overlap distributions above and below the
spin-glass transition have a direct analogies to population structure and
heterozygosity: In the high recombination limit, genotypes in the population are
assembled from the available alleles more or less at random such that any two
individuals differ at about $2\sum_{i=1}^L \nu_i(1-\nu_i)$ sites ($\nu_i$ being
the allele frequency at locus $i$). At low recombination (or substantial inbreeding),
the population will condense into fit genotypes (or inbred groups) that are much
more similar to each other than to members of the general population.

Figure \ref{fig:phasediagram}B illustrates pair-wise interaction between
polymorphic loci along the chromosome. In general, we expect a complex and
possibly hierarchical pattern of interactions: A given pair of distant genes
will have only a low probability to interact substantially, while polymorphisms
within one gene and its regulatory elements are much more likely to interact strongly. 
Nearby polymorphisms in a protein \citep{Callahan_2010} will be still more likely to interact. 
In obligate sexuals, the sparse long range interactions will rarely
suffice to produce appreciable correlations between loci. Within small stretches
of chromosomes, however, recombination rates are low and if the strength of
interactions within this stretch is sufficiently high, QLE will locally break
down. Consider for example a one centimorgan long region, which in humans corresponds to
about one mega base and harbors around a thousand polymorphisms. If the typical
epistatic contribution fitness of this stretch of chromosome were on the order of
1\%, we expect run-away selection on coadapted haplotypes and strong
correlations. Since distant parts of the genome are in QLE, one expects a
``module'' selection regime, where loosely linked and weakly interacting
``modules'' are in QLE, but strong interactions and infrequent recombination has
led locally to condensation into coadapted haplotypes \citep{Neher:2009p22302}.
(An excellent early discussion of such epistasis driven ``coagulation" in the 
``soup" of genes is found in \citep{Turner:1967p11866}.)
Put otherwise, we can view such a system as consisting of weakly-interacting
mesoscopic loci, at which several super-alleles segregate. These super-alleles
are ``destructible'', in the sense that recombination within leads to reduced
fitness and purging by selection. However, since recombination within these
alleles is rare, quantitative traits would be highly heritable on short
timescales and quantitative genetics would work as usual. 

Our discussion of the multilocus theory and QLE  was guided by ideas of
Statistical Physics. The explicit form of the (approximate) genotype distribution
function $P(\gt)$ parametrized by instantaneous allele frequencies is the central
pillar connecting the dynamics of population average traits -  the subject of QG
- to the individual-based evolutionary process. It is essential that the QLE
distribution is reached on a relatively fast time scale of mating and
recombination. Allele frequencies are well defined and vary slowly on this time
scale. The QLE ensemble should not be confused with a very different
mutation/selection/drift ensemble - which could be rightfully termed the ``Wright
equilibrium" (Eq. (\ref{eq:wright equilibrium})) - which is often invoked as a
link between evolutionary dynamics and Statistical Physics. Wright equilibrium
gives a stationary distribution of allele frequencies which would be established
in a finite population ($N$ playing the role of inverse temperature) on a time
scale longer than the inverse mutation rate $\mu$, provided stationary selection
pressures. Ruggedness of the fitness landscape could further increase this
equilibration time scale exponentially \citep{Weissman:2010p37077}. Clearly, this
type of equilibrium applies on a very different time scale than the phenomena
addressed in the present work. Related ideas were developed in the context of
quasi-species theory to study the conditions under which hereditary information
can be maintained over long times \citep{Eigen:1971p11019,Franz:1997p1708}. The
focus of these studies was pre-biotic evolution, where fidelity of replication
was most likely low and stability of genomic information can be sensibly studied
using a simple equilibrium model. Equilibrium arguments were also applied to
evolution of codon bias \citep{Iwasa:1988p29290} and the evolution of transcription factor
binding sites \citep{Mustonen:2005p13413}. In the latter two cases, an ensemble
can be constructed by combining many instances of the same sequence motive which
was under constant selection pressure for very long time (conserved transcription
factor binding motive or conserved preference of certain codons over others). In
many cases, however, the equilibrium state is of little relevance.


In conclusion, in this review we have provided a derivation of the genotype
distribution in the QLE approximation, providing a systematic generalization of
Fisher's theorem, Kimura's diffusion theory and Wright's equilibrium from
LE to QLE, which includes the effect of (weak) correlations between loci. We
have also discussed the limitation of the QLE approximation and the structure of
the genotype distribution at low recombination rates. 

It is our hope that better understanding of the QLE approximation  will promote
progress in understanding the effects associated with its breakdown, whether due
to strong epistasis or strong physical linkage, such as for example the
Hill-Roberson effects (hitch-hiking and background selection) which still await
comprehensive treatment.

\begin{acknowledgments}
The authors have benefitted from interaction with
many colleagues including N.~Barton, M.~Desai, D.S.~Fisher, A.~Dayarian,
S.~Goyal, M.~Kreitman, M.~Lynch, P.~Neveu, E.~Siggia, H.~Teotonio and M.~Vergassola. This work
was supported by the National Science Foundation under grant no.~PHY05-51164 (RAN), PHY-0844989 (BIS) and by the Harvey L. Karp Discovery Award (to RAN).
\end{acknowledgments}

\appendix
\section{Glossary}
\label{sec:glossary}
\begin{description}
\item[Allele] State of a locus, for example the base \texttt{A}, \texttt{C},
\texttt{G} or \texttt{T} at a certain position
\item[Crossover rate] In meiosis, parental chromosomes are paired up and
crossed over. The density of crossovers on the chromosome is called
crossover rate.
\item[Dominance] Interaction of the two alleles at the same locus in diploid
organisms
\item[Epistasis] Genetic interactions between alleles at different loci,
i.e.~a dependence of the effect of an allele at one locus on the remainder of
the genome.
\item[Fitness] Expected reproductive success of an organism. For modeling
purposes, this is often equated with the growth rate (Malthusian or log fitness)
or the average number of offspring in the subsequent generation (absolute fitness).
\item[Gametes] Egg and sperm
\item[Genetic Drift] Sampling fluctuations of genotype or allele frequencies.
Genetic drift enters as the diffusion term in the Fokker-Planck equation
for the dynamics of the distribution of allele frequencies.
\item[Genetic map] The cumulative crossover rate along the chromosome. The
average number of crossover events per chromosome is the map length.
\item[Genotype] State of the genome, i.e.~the set of alleles an individual
carries.
\item[Haplotype] Alleles inherited from one parent. In diploids, two haplotypes
make one genotype.
\item[Heritability] Broad sense heritability is the genetic component of
traits, i.e.~the concordance of traits between monozygotic twins. Narrow
sense heritability refers to the genetic component of traits that is inherited
in sexual reproduction, i.e.~the correlation between trait values of parents
and children.
\item[Heterozygosity] Fraction individuals in a diploid
population that carry distinct alleles at a locus.
\item[Homozygosity] The complement of heterozygosity.
\item[Linkage] Loci on the same chromosome are linked and share history until
crossover events separate them.
\item[Linkage (dis)equilibrium] Absence (presence) of correlations between loci
\item[Locus] Location on the chromosome, e.g.~a gene.
\item[Mean Fitness] To preserve overall population size, fitness is often
measured with respect to the mean fitness of the population.
\item[Meiosis] Division of a diploid cell to produce haploid gametes
\item[Panmictic] A population is panmictic if each individual is equally
likely to compete and interact with any other individual. In practice, this
requires that dispersal is fast compared to population genetic time scales. 
\item[Polymorphism] A locus with variation, i.e.~the population contains several
alleles at this locus.
\item[Random mating] Simplifying assumption that mating is independent of
genotype, phenotype, and environment.
\item[Recombination] Process of reshuffling of the genetic material in sexual
reproduction.
\item[Outcrossing] Fertilization with sperm/pollen from a different individual 
\item[Selfing] Many plants and other organisms have female
and male sexual organs and can self-fertilize or self-pollinate.
\end{description}

\section{Notation}
\label{sec:notation}
The specification of genotypes and parameterization of genotype-phenotype maps is
not unique and our notation differs from the traditional population genetics
choice. Conventionally, one chooses one ``wild-type'' reference genome
$(0,0,\ldots, 0)$ and enumerates deviations from this reference. This is useful
when a well defined wild-type genotype exists. In diverse populations, for
example the progeny of cross between diverged strains, the reference free
parameterization we are using here is more natural. The allelic state at each
locus is denoted symmetrically by $\pm 1$, e.g.~whether an allele comes from one
or the other strain. The two different parameterizations are completely
equivalent and related to each other by a simple linear transformation (see
Table \ref{tab:translation} below). In the present context the reference free notation
simplifies the algebra since the $s_i = \pm 1$ basis is orthogonal when averaging over the genotype
space. The relation to the Fourier transform allows an unambiguous decomposition
of the fitness function into additive parts and epistatic components of different
order (Parceval's Theorem), while in the reference based parametrization of
fitness functions, more akin to a Taylor expansion, coefficients depend
explicitly on the choice of reference. We have also deviated from the traditional
$D_{ij}$ notation for linkage disequilibrium because we want to use the diagonal
$\chi_{ii}=1-\chi_i^2$ components of the cumulant matrix (two times the
heterozygosity at locus $i$) on the same footing as the off-diagonal one.

\begin{table}[h]
\begin{tabular}{ll}
\hline
{\bf Symbol} & {\bf Meaning/Definition} \\ \hline
$\gt$ & Haploid genotype: $\gt=\{\locus_1, \ldots, \locus_L\}$\\
$P(\gt, t)$ & Genotype distribution in the population\\
$\langle ... \rangle$ & Population average\\
$\fit(\gt)$ & Fitness (growth rate) of genotype $\gt$\\
$\fitcoeff_{i_1\ldots i_k}$ & Contribution to fitness of the $i_1\ldots i_k$ set
of loci\\ 
$\acoeff_i$ & Additive effect of locus $i$\\
$\sigma^2$, $\sigA$, $\sigI$ & Total, additive, and epistatic variance in
fitness\\
$\xi_i\in\{0,1\}$ & Origin of locus $i$, i.e.~maternal or paternal\\ 
$C(\{\xi_i\})$ & Probability of the recombination pattern
$\{ \xi_i\}$\\
$\xo_{ij}$ & Probability that  loci $i$ and $j$ derive \\ &from different
parents\\ $\mu$, $\rec$ & Mutation and outcrossing rate \\ 
\hline
\end{tabular}
\caption{Table of symbols}
\label{tab:symbols}
\end{table}
\begin{table}[h]
\begin{tabular}{l|l}
\hline
{\bf Quantity} &{\bf Our notation}  \\ \hline
Allele at locus $i$, $\{ a_i, A_i\}$ &$\locus_i\in \{-1, 1\}$ \\
Allele frequency $\nu_i$ & $\nu_i= (1+\chi_i)/2$ where $\chi_i = \langle
\locus_i \rangle$ \\ 
Linkage Disequilibrium  $D_{ij}$ ($i\neq j$) &
$4D_{ij}=\chi_{ij}=\langle \locus_i \locus_j \rangle-\langle \locus_i 
\rangle \langle  \locus_j \rangle $\\ Heterozygocity $H_i=2\nu_i (1-\nu_i)$ &
$2 H_i=\chi_{ii}=(1-\chi_i^2)$ \\
\hline
\hline
\end{tabular}
\caption{Population genetic quantities in our notation}
\label{tab:translation}
\end{table}

\section{QLE in terms of effective fields}
\label{sec:app_QLE}
In this appendix, we discuss how the fields $\phi_i$ and $\phi_{ij}$ introduced
to parameterize the genotype distribution $P(\gt,t)$ in Eq.~(\ref{eq:logP}) are
related to the cumulants of $P(\gt,t)$. We also detail how the recombination
term in Eq.~(\ref{eq:genotype_dynamics}) can be evaluated explicitly within
the QLE perturbation theory. We parameterized
the genotype distribution via
\begin{equation}
\log P(\gt , t) =  {\Phi}(t) +\sum_i \phi_i (t) \locus_i + \sum_{i<j}
\phi_{ij} (t) \locus_i \locus_j.
\end{equation}
The constant term is determined by the normalization of the distribution, 
the coefficients $\phi_i(t)$ are related to frequencies and the second order
coefficients $\phi_{ij}(t)$ to the connected correlation between loci. 
In the limit under consideration, the second order contributions are small and
we evaluate the coefficients to leading order in
$\phi_{ij}(t)$.

\begin{equation}
\begin{split}
e^{-{\Phi}} & = \sum_{\gt} e^{\sum_i \phi_i  \locus_i + \sum_{i<j}
\phi_{ij}\locus_i\locus_j} \\ &\approx \sum_{\gt} e^{\sum_i \phi_i 
\locus_i }\left(1 +\sum_{i<j}
\phi_{ij}\locus_i\locus_j\right) 
\\ &= 2^L\left(1+\sum_{k<j} \phi_{kj}
\tanh(\phi_k)\tanh(\phi_j)\right)\prod_{i=1}^L \cosh(\phi_i)
\end{split}
\end{equation}
The relations between $\chi_i$, $\chi_{ij}$ and $\phi_i$, $\phi_{ij}$
given in Eq.~(\ref{eq:chi_to_phi}) follow by differentiation.

To arrive at the equations for the time evolution of the fields $\phi_i$ and
$\phi_{ij}$ (Eq.~(\ref{eq:d_t_phi})), we have to evaluate the recombination term
in Eq.~(\ref{eq:logPdt}). This is done below. The terms proportional to $\phi_i(t)$ cancel exactly between
numerator and denominator and we are left with
\begin{equation}
\label{eq:rec_term}
\begin{split}
\sum_{\{\xi_i\}\{\locus'_i\}}
& C(\{\xi\})P(\gt') \left[\frac{P(\gtm)P(\gtf)}{P(\gt)P(\gt')}-1\right] \\
&=\rec\sum_{\{\xi_i\}\{\locus'_i\}} C(\{\xi\})P(\gt')\left[e^{\sum_{i<j}
\phi_{ij}\left[ (\xi_i \locus_i+\bar{\xi}_i\locus_i')(\xi_j
\locus_j+\bar{\xi}_j\locus_j')+(\bar{\xi}_i
\locus_i+\xi_i\locus_i')(\bar{\xi}_i
\locus_i+\xi_j\locus_j') - \locus_i\locus_j - \locus_i'\locus_j'\right]}
-1\right] \\
&= \rec \sum_{\{\xi_i\}\{\locus'_i\}}
C(\{\xi\})P(\gt')\left[e^{\sum_{i<j} \phi_{ij}\left[ (\xi_i \xi_j +
\bar{\xi}_i\bar{\xi}_j -1 )(\locus_i \locus_j+\locus_i'
\locus_j') +  (\xi_i \bar{\xi}_j +
\bar{\xi}_i \xi_j)(\locus_i \locus_j'+\locus_i' \locus_j)\right]}-1\right]
\end{split}
\end{equation}
In the limit under consideration, the second order contributions have to be
small enough that the entire exponent is small. In this case, the exponential
can be expanded and the different terms averaged individually.
\begin{equation}
\label{eq:rec_term_digest}
\begin{split}
\sum_{\{\xi_i\}\{\locus'_i\}}
& C(\{\xi\})P(\gt') \left[\frac{P(\gtm)P(\gtf)}{P(\gt)P(\gt')}-1\right] \\
&\approx \rec \sum_{\{\xi_i\}}
C(\{\xi\})\sum_{i<j} \phi_{ij}\left[ (\xi_i \xi_j +
\bar{\xi}_i\bar{\xi}_j -1 )(\locus_i \locus_j+\langle \locus_i\locus_j\rangle)
+ (\xi_i \bar{\xi}_j + \bar{\xi}_i \xi_j)(\locus_i\langle
\locus_j\rangle+\langle \locus_i\rangle \locus_j)\right]\\
&=  \sum_{i<j} \xo_{ij}\phi_{ij}\left[(\locus_i\langle
\locus_j\rangle+\langle \locus_i\rangle \locus_j)- (\locus_i \locus_j+\langle \locus_i\locus_j\rangle) \right]
\end{split}
\end{equation}
where $\xo_{ij}$ is the probability that an odd number of crossovers happened
between loci $i$ and $j$.

\section{Diffusion theory and Wright's equilibrium}
\label{sec:app_wright}
In this appendix, we detail intermediate steps to arrive at the diffusion
equation for the allele frequencies in QLE and the generalized Wright equilibrium. 
The noise terms in the Langevin equation (\ref{eq:langevin}) stem from the
multinomial sampling of the genotypes or gametes. From the covariance of the
multinomial distribution, we can therefore determine the covariance of the
noise terms $\zeta_i$ for the $\chi_i$ and the $\zeta_{ij}$ for the $\chi_{ij}$.
The covariance the changes in $\chi_i$ and $\chi_j$, for example, can be
calculated as follows
\begin{equation}
\begin{split}
\langle \Delta\chi_i \Delta\chi_j \rangle &= \langle \sum_{\gt}\locus_i
\Delta P(\gt)\sum_{\gt}\locus_j' \Delta P(\gt')\rangle  =  \sum_{\gt
\gt'}\locus_i\locus_j'   \langle\Delta P(\gt)\Delta P(\gt')\rangle \\
&=\frac{1}{N}\left[\sum_{\gt}\locus_i\locus_j  P(\gt)(1-P(\gt)) - \sum_{\gt\neq
\gt'}\locus_i\locus_j'   P(\gt)P(\gt') \right] = \frac{\chi_{ij}}{N}
\end{split}
\end{equation}
The other covariance terms can be calculated analogously. In particular, one
finds $\langle \Delta\chi_{ij}^2 \rangle \approx N^{-1}\chi_{ii}\chi_{jj}=
N^{-1}(1-\chi_{i}^2)(1-\chi_{j}^2)$.

\subsection*{The effect of deterministic correlations}
If deterministic correlations dominate over the fluctuations in $\chi_{ij}$ the
forward Kolmogorov equation for the distribution of the $\chi_i$ is given by
\begin{equation}
\label{eq:FPE}
\partial_t Q(\{\chi_i\},t) = \sum_i \partial_{\chi_i} \left[
\frac{1}{2N}\sum_{j} \partial_{\chi_j} (\chi_{ij}Q(\{\chi_i\},t) ) +
Q(\{\chi_i\},t) \left ( 2\mu\chi_i-\sum_j \chi_{ij}\partial_{\chi_j} \mfit \right )\right] \ .
\end{equation}
In the steady state, all probability fluxes vanish. The $i$ component of the
probability flux is precisely the expression in brackets above and hence has to be equal to zero. Multiplying the bracket
with $2N\chi_{ki}^{-1}$ and summing over $i$ ($\chi_{ki}^{-1}$ is the $ki$ element of the matrix inverse of
$\chi_{ij}$), we have
\begin{equation}
\label{eq:SUPsteadystate}
\partial_k Q(\{\chi_i\}) =  Q(\{\chi_i\}) \left (-\sum_{ij}\chi_{ki}^{-1}
\partial_j \chi_{ij}- 4N\mu\sum_i \chi_{ki}^{-1}\chi_i+ \partial_{\chi_k} \mfit \right )
\end{equation}
Next, we use the fact that the off-diagonal elements of $\chi_{ij}$ are small
and $\chi_{ij} = \gamma_{ij}(1-\chi_i^2)(1-\chi_j^2)$, while
$\chi_{ii}=1-\chi_i^2$. To first order in the off-diagonal elements, the inverse is given by
$\chi_{ii}^{-1} = (1-\chi_i^2)^{-1}$ and off-diagonal elements $\chi_{ij}^{-1}
=-\chi_{ij}(1-\chi_j^2)^{-1}(1-\chi_i^2)^{-1}=-\gamma_{ij}$. Going over the
terms in Eq.~(\ref{eq:SUPsteadystate}) one by one, we have
\begin{equation}
\begin{split}
\sum_{ij}\chi_{ki}^{-1} \partial_{\chi_j} \chi_{ij} &= \sum_i \chi_{ki}^{-1}
\sum_{j\neq i} \partial_{\chi_j} \chi_{ij}  + \sum_i \chi_{ki}^{-1}
\partial_{\chi_i} \chi_{ii} \\
&=\sum_i \chi_{ki}^{-1} \chi_{ij}\sum_{j\neq i} \partial_{\chi_j} \log(1-\chi_j^2)   +
\sum_i
\chi_{ki}^{-1} \chi_{ii}\partial_{\chi_i} \log(1-\chi_i^2)  \\
&=\partial_{\chi_k} \log(1-\chi_k^2) 
\end{split}
\end{equation}
The mutation term can be evaluated as follows
\begin{equation}
\begin{split}
4N\mu\sum_i \chi_{ki}^{-1}\chi_i &=- 2N\mu \partial_{\chi_k} \log(1-\chi_k^2) -
4N\mu \sum_{i\neq k} \gamma_{ij} \chi_i \\
&=- 2N\mu \partial_{\chi_k} \log(1-\chi_k^2) - 4N\mu
\partial_{\chi_k}\sum_{i\neq k} \gamma_{ik} \chi_i\chi_k
\end{split}
\end{equation}
Substituting these terms into Eq.~(\ref{eq:SUPsteadystate}) and $\gamma_{ij} =
\frac{\fitcoeff_{ij}}{\rec\xo_{ij}}$, we have
\begin{equation}
\partial_k Q(\{\chi_i\}) =  Q(\{\chi_i\}) \partial_{\chi_k}\left
((2N\mu-1)\log(1-\chi_k^2) + 2N[\mfit + 2\mu\sum_{i\neq k} \frac{\fitcoeff_{ik}\chi_i\chi_k}{\rec\xo_{ik}}] \right)
\end{equation}
which is straight-fowardly integrated to 
\begin{equation}
Q(\{\chi_i\}) = C e^{2N\mfit + 4N\mu\sum_{i<k}
\frac{\fitcoeff_{ik}\chi_i\chi_k}{\rec\xo_{ik}}}
\prod_{i=1}^L(1-\chi_i^2)^{2N\mu-1}
\end{equation}

\subsection*{The effect of fluctuating correlations between loci}
Even when associations between loci are zero on average, fluctuations of
$\chi_{ij}$ can affect the allele frequency dynamics. The coupling between
different loci acts as an additional noise source on the dynamics of
allele frequencies. Grouping deterministic and stochastic forces, the
corresponding Langevin equation for $\chi_i$ is given by
\begin{equation}
\label{eq:appNeLangevin}
\chi_i(t+\Delta t)-\chi_i(t) = \Delta
t\left[\chi_{ii}\partial_{\chi_i}\mfit-2\mu\chi_i\right]  + \int_{t}^{t+\Delta
t} dt'\left[\sum_{j\neq i}\chi_{ij}(t') \partial_{\chi_j} \mfit + \zeta_i\right]
\end{equation}
where the integral constitutes the fluctuating noise term. Solving the Langevin
equation for $\chi_{ij}(t)$ assuming constant $\chi_i$ and $\chi_j$, one finds
\begin{equation}
\langle \chi_{ij}(t)\chi_{ij}(t+\Delta t)\rangle =
\frac{(1-\chi_i^2)(1-\chi_j^2)e^{-\rec\xo_{ij}\Delta t}}{2Nr\xo_{ij}}
\end{equation}
Averaging the square of the noise term in Eq.~(\ref{eq:appNeLangevin}), we find
\begin{equation}
\begin{split}
&\langle \int_t^{t+\Delta t}dt'\int_t^{t+\Delta t}dt'' \left[\sum_{j\neq
i}\chi_{ij} \partial_{\chi_j} \mfit + \zeta_i\right]\left[\sum_{k\neq
i}\chi_{ik} \partial_{\chi_j} \mfit + \zeta_i\right]\rangle \\
&\approx \frac{\chi_{ii}\Delta T}{N}\left[ 1+  \frac{1}{2}\sum_{j\neq
i}\chi_{jj} \left(\frac{1}{\rec \xo_{ij}}\frac{\partial
\mfit}{\partial \chi_j}\right)^2\right]
\end{split}
\end{equation}
The cross-term is of order $N^{-3/2}$ and can be neglected.

\bibliography{bib_nourl}
\end{document}